\documentclass[fleqn,usenatbib]{mnras}

\usepackage{newtxtext,newtxmath}
\usepackage{rotating}  % Include the rotating package for sidewaystable
\usepackage{afterpage}
\usepackage{pdflscape}
\usepackage[T1]{fontenc}

\DeclareRobustCommand{\VAN}[3]{#2}
\let\VANthebibliography\thebibliography
\def\thebibliography{\DeclareRobustCommand{\VAN}[3]{##3}\VANthebibliography}

\usepackage{graphicx}	% Including figure files
\usepackage{amsmath}	% Advanced maths commands
\usepackage{threeparttable}
\title[{\tt DOLBY} method of radial-velocity extraction]{BEBOP VI. Enabling the detection of circumbinary planets orbiting double-lined binaries with the {\tt DOLBY} method of radial-velocity extraction}

\author[S. Lalitha  et al.]{
Lalitha Sairam,$^{1,2}$\thanks{E-mail: ls2071@cam.ac.uk}
Thomas A. Baycroft,$^{2}$
Isabelle Boisse,$^{3}$
Neda Heidari,$^{4}$
Alexandre Santerne,$^{3}$\newauthor
Amaury H.M.J. Triaud,$^{2}$
Gavin A.L. Coleman,$^{5}$
Yasmin T. Davis,$^{2}$
Magali Deleuil,$^{3}$
Guillaume H\'ebrard,$^{4}$\newauthor
David V. Martin,$^{6}$
Pierre F.L. Maxted,$^{7}$
Richard P. Nelson,$^{5}$
Daniel Sebastian,$^{2}$
Owen J. Scutt,$^{2}$\newauthor
Matthew R. Standing$^{8}$ \\
$^{1}$Institute of Astronomy, University of Cambridge, Madingley road, Cambridge CB3 0HA, UK\\
$^{2}$School of Physics and Astronomy, University of Birmingham, Edgbaston,Birmingham B15 2TT, UK\\
$^{3}$Aix Marseille Univ, CNRS, CNES, LAM, Marseille, France\\
$^{4}$Institut d'astrophysique de Paris, UMR 7095 CNRS université pierre et
marie curie, 98 bis, boulevard Arago,  75014, Paris\\
$^{5}$Astronomy Unit, Queen Mary University of London, Mile End Road, London E1 4NS, UK\\
$^{6}$Department of Physics and Astronomy, Tufts University, 574 Boston Avenue, Medford, MA 02155\\
$^{7}$ Astrophysics Group, Keele University, ST5 5BG, UK\\
$^{8}$ European Space Agency (ESA), European Space Astronomy Centre (ESAC), Madrid, Spain\\
}

\date{Accepted XXX. Received YYY; in original form ZZZ}

\pubyear{2024}

\begin{document}
\label{firstpage}
\pagerange{\pageref{firstpage}--\pageref{lastpage}}
\maketitle

\begin{abstract}

%Because they orbit about both stars of a binary, circumbinary planets offer an opportunity to study planet formation and planet orbital migration in a slightly altered environment compared to a single star's. 
Circumbinary planets - planets that orbit both stars in a binary system - offer the opportunity to study planet formation and orbital migration in a different environment compare to single stars. 
However, despite the fact that $> 90\%$ of binary systems in the solar neighbourhood are  spectrally resolved double-lined binaries, there has been only one detection of a circumbinary planet orbitting a double-lined binary using the radial velocity method so far. Spectrally disentangling both components of a binary system is hard to do accurately. Weak spectral lines blend with one another in a time-varying way, and inaccuracy in spectral modelling can lead to an inaccurate estimation of the radial-velocity of each component. This inaccuracy adds scatter to the measurements that can hide the weak radial-velocity signature of circumbinary exoplanets. We have obtained new high signal-to-noise and high-resolution spectra with the SOPHIE spectrograph, mounted on the 193cm telescope at Observatoire de Haute-Provence (OHP) for six, bright, double-lined binaries for which a circumbinary exoplanet detection has been attempted in the past. To extract radial-velocities we use the {\tt DOLBY} code, a recent method of spectral disentangling using Gaussian processes to model the time-varying components. We analyse the resulting radial-velocities with a diffusive nested sampler to seek planets, and compute sensitivity limits.\\ We do not detect any new circumbinary planet. However, we show that the combination of new data, new radial-velocity extraction methods, and new statistical methods to determine a dataset's sensitivity to planets leads to an approximately one order of magnitude improvement compared to previous results. This improvement brings us into the range of known circumbinary exoplanets and paves the way for new campaigns of observations targeting double-lined binaries.
\end{abstract}


%%%%%%%%%%%%%%%%%%%%%%%%%%%%%%%%%%%%%%%%%%%%%%%%%%

%%%%%%%%%%%%%%%%% BODY OF PAPER %%%%%%%%%%%%%%%%%%

\section{Introduction}

One of the most remarkable discoveries from the \emph{Kepler} mission is the identification of the first confirmed circumbinary planet, Kepler-16\,b \citep{Doyle_2011}. This planet orbits both stars of its binary host providing a novel environment to investigate the outcome of planet formation. The discovery of Kepler-16\,b was surprising because theoretical work implies that circumbinary configurations prevent the formation of planets at such proximity to the binary \citep[e.g.][]{Meschiari_2012, Paardekooper_2012, Lines2014, Pierens2020, Martin2022}. In parallel, once planets have formed, disc-driven orbital migration parks the planets near the instability limit \citep[e.g.][]{Dvorak1989, holman_1999,doolin2011, Coleman2023, Coleman2024} in ways that encodes the conditions of their protoplanetary disc into their observable orbital properties \citep[e.g.][]{Penzlin2021}. Thus, circumbinary planets offer new insights into planet formation mechanisms that will inform planet formation around single stars and that enhance our understanding of the exoplanet population at large, completing the tale behind the origins of the Solar System, and of our Earth.

Few circumbinary planets have been discovered so far, and most of them were found by analysing months-long space-based photometric time-series. This is expensive, and inefficient. A ground-based radial velocity survey of binaries is expected to be less biased towards planetary orbital plane perpendicular to the sky than the transit method, while finding as many planets in only a fraction of the telescope time, measuring more orbital parameters, and constructing a leaner, more intuitive picture of the circumbinary population.  At the moment, single-lined binaries (SB1) are the only binaries for which a 1-2 $\rm\,m\,s^{-1}$ radial velocity accuracy has been demonstrated \citep{standing_2022, Triaud2022}, thus allowing for a more reliable planet detection success rate. The BEBOP survey \citep[Binary Escorted by Orbiting Planets;][]{Martin2019} has monitored a sample of single-lined binary systems in search of circumbinary planets leading to planet detections \citep[e.g. TOI 1338c/ BEBOP-1c;][]{Standing2023}. Several other candidates have also been announced (Baycroft et al. in press). However, in order to truly understand circumbinary systems and their formation, we need to expand the type of binaries around which circumbinary planets can be detected.

Around $90\%$ of binary systems in the solar-neighbourhood are spectrally resolved as double-lined binaries \citep[SB2;][]{Kovaleva_2016}, possibly making them the most common circumbinary planet-hosts. This is an opportunity to dramatically increase the sample available for circumbinary planet surveys, and perform better comparisons between the properties of exoplanets orbiting singles stars, and the properties of circumbinary planets. Double-lined binaries are on average brighter, further helping with the detection of small-mass exoplanets orbiting them. Furthermore, for each spectrum, two radial velocity measurements are obtained (for each of the components).
Additionally, the vast majority of NASA's {\it TESS} mission \citep{Ricker2014} might produce serendipitous circumbinary planet candidates as a "{\it 1--2 punch}" transit event on double-lined binaries \citep{Kostov2020}. For most of the sky, only 27-day timeseries are available in {\it TESS}, and are themselves obtained once typically every other year. Because of circumbinary planet's intrinsically long orbital periods, this implies two transits of a circumbinary planet transits roughly one orbital period away from one another are challenging to obtain \citep[except in {\it TESS}'s continuous viewing zone as in][]{Kostov2021}. However, within those short sectors, a circumbinary planet can still be identified when it first transits one star, and soon after transits the other star of the binary system, a {\it 1--2 punch}. For both transit events to be detectable, both stars need to be of comparable brightness, hence, they will most likely be double-lined binaries. Without at least a 1-2 $\rm\,m\,s^{-1}$ accuracy on double-lined binaries, we might never known whether those events are truly produced by planets.

A dedicated search for circumbinary planets using the radial velocity method, TATOOINE, was attempted on ten double-lined binaries \citep[SB2;][]{Konacki_2009, Konacki2010}. Despite being the state-of-the-art, their results showed a large scatter of 10-15$~\rm m\,s^{-1}$ in radial velocities (sometimes more), although photon noise reaches $1~\rm m\,s^{-1}$ in some cases (they also had to deconvolve an iodine reference spectrum). This extra scatter would prevent the detection of most circumbinary gas-giants (assuming similar planet population properties between single and binary stars). This 10-15$~\rm m\,s^{-1}$ problem is likely caused by an imperfect disentangling of the time-varying blending of weak spectral lines, as the two components orbit one another, translating across the spectrograph by $\sim100\,\rm km\,s^{-1}$. \cite{Konacki_2009} disentangled their spectra to achieve a remarkable precision even at current standards, but it was not enough for identifying planets. Many lines have small signal-to-noise on each spectrum and are easy to miss. While individually they appear as noise, when hundreds of those are missed, they likely contribute to a spurious signal. Binary mask cross-correlation methods also handle line blends badly and those regions are usually avoided. They too produce extra-noise on double-lined binaries, even in 2-dimensional cross-correlations \citep{Zucker_2004}.

As part of the BEBOP programme, we invested a small amount of telescope time to observing double-lined binaries, and in creating new ways of extracting radial-velocities for SB2s. In \cite{lalitha_2023}, we produced two novel methods (now called {\tt DOLBY}; DOuble-Lined BinarY) able to precisely measure the radial velocities of double-lined binaries. {\tt DOLBY} uses a Gaussian process to help model both spectral components and disentangle them from one another. Applying the {\tt DOLBY} methods to TIC~172900988, a proposed circumbinary planet transiting a double-lined binary \citep{Kostov2021}, we successfully detected a Doppler variation consistent with a circumbinary planet at an orbital period of $\sim 150\rm \, d$ \citep{lalitha_2023}. However, the stars composing the TIC~172900988 system are both fairly hot, for the primary star, we obtained radial-velocity scatters of approximately 40 and 49 $\rm m\,s^{-1}$ using both the {\tt DOLBY}-SD and {\tt DOLBY}-CCF methods, and for the secondary, we obtained scatters of approximately 50 and 72 $\rm m\,s^{-1}$, respectively. This scatter did not fully demonstrate the potential of our new approach. In the current paper, we report the {\tt DOLBY} analysis of our observations of another six SB2 systems that were collected with the SOPHIE spectrograph \citep{Perruchot_2008}. All six were specifically taken from the studies by \citet{Konacki_2009} and \citet{Konacki2010} to enable a comparison with their results, and test how well the {\tt DOLBY} methods perform.

This paper is structured as follows: \S\ref{sec:obs} details the observations of SB2 systems used in our analysis. \S\ref{sec:method} describes the method employed to measure precise radial velocities for the binaries. In \S\ref{sec:results}, we delve into the analysis of the radial velocities, including a comparison between those derived from the traditional TODMOR method. We also discuss the detection limit achieved for the circumbinary method using {\tt kima} for each target system. We conclude the paper with a summary of our key results in \S\ref{sec:conclusion}.

\section{A summary of our double-lined binary observations}\label{sec:obs}

In 2019, we initiated a sub-programme targeting 10 double-lined binaries with the SOPHIE spectrograph mounted on the 193cm at Observatoire de Haute-Provence \citep{Perruchot_2008}. Our primary objective was to collect a representative sample of double-line binaries, bright and not so bright, across various spectral types and different binary orbital periods with the goal of developing and testing new methods of radial-velocity extraction. Our secondary objective was to detect circumbinary planets in these systems. Our full list includes TIC~172900988, where we first presented the {\tt DOLBY} methods and detected the circumbinary planet TIC~172900988\,b \citep{lalitha_2023}. We also survey another three double-lined binaries known to host circumbinary planets, from the {\it Kepler} field. These are typically faint. The remaining six systems, the subjects of this paper, are HD\,195987, HD\,210027, HD\,9939, HD\,78418, HD\,13974, and HD\,282975 \citep[from the TATOOINE survey;][]{Konacki_2009,Konacki2010}. Their properties range $3.76<{\rm Vmag}<10$, F$5<{\rm type}<$G6 and $10<{\rm P_{bin}}<57~\rm d$. Compared to the median ${\rm Vmag} \sim 11.5$ from BEBOP's main sample of single-lined eclipsing binaries \citep{Martin2019}, these six systems are much brighter and highlight well our motivation for observing them and improving radial-velocity measurements for SB2s. The fundamental stellar properties of the targets are summarised in Table \ref{tab:tab1}.

Between 2019-10-09 and 2023-05-06 we collected 244 spectra on the six binaries, with exposure times ranging from 200 to 1800s. We obtained a median signal-to-noise ratio, ${\rm SNR}\approx 85$ at $5500$ \AA\,. 
%SNR for HD9939 67.21, HD195987 68.88, HD210027 189.55, HD282975 38.12, HD78418 101.66, HD13974 119.49
Some of the observations were obtained in adverse weather conditions, since those systems are bright and could serve as back-up targets. Observations are summarised in Table~\ref{tab:tab1}, and spectra are publicly available in the SOPHIE archive.

SOPHIE's high-resolution spectra cover a wide wavelength range from $3872$ to $6943$ \AA\, divided into 39 spectral orders, with a resolving power of $\lambda/ \delta\lambda \approx75,000$. The SOPHIE spectrograph was specifically designed to achieve long-term stability of $1-2\,\rm m\,s^{-1}$, enabling the detection of exoplanets. During the observations, we employed the {\tt fpsimult} mode, utilising one fibre to capture starlight and another fibre towards a Fabry-P\'erot lamp, to measure the instrument drift. %to calibrate the wavelength solution.
To ensure accurate wavelength calibration, we performed calibration procedures before the start of each night using a thorium-argon lamp and a Fabry-P\'erot.

The data reduction process involved the SOPHIE automatic pipeline \citep{Bouchy_2009b}, which handled the extraction of the spectra and performed the necessary wavelength calibration. As described in \cite{heidari2024sophie}, an additional colour correction was applied to account for instrumental effects.  To ensure the accuracy of our spectroscopic data, we also applied a charge transfer inefficiency (CTI) correction to the SOPHIE spectra. The CTI correction\footnote{For HD9939 and HD195987 the CTI correction appears to induce a large residual scatter in the data for the secondary star, so for these two cases we do not apply the correction} was performed using the method described in \citet{Bouchy_2009b}. Subsequently, we cross-correlated the resulting wavelength-calibrated CTI corrected spectra with a numerical binary mask \citep{Baranne_1996, Pepe_2002}. This cross-correlation step allowed us to extract the cross-correlation functions (CCFs), which are essential for our subsequent analysis.

\begin{figure*}
\includegraphics[width=0.98\textwidth]{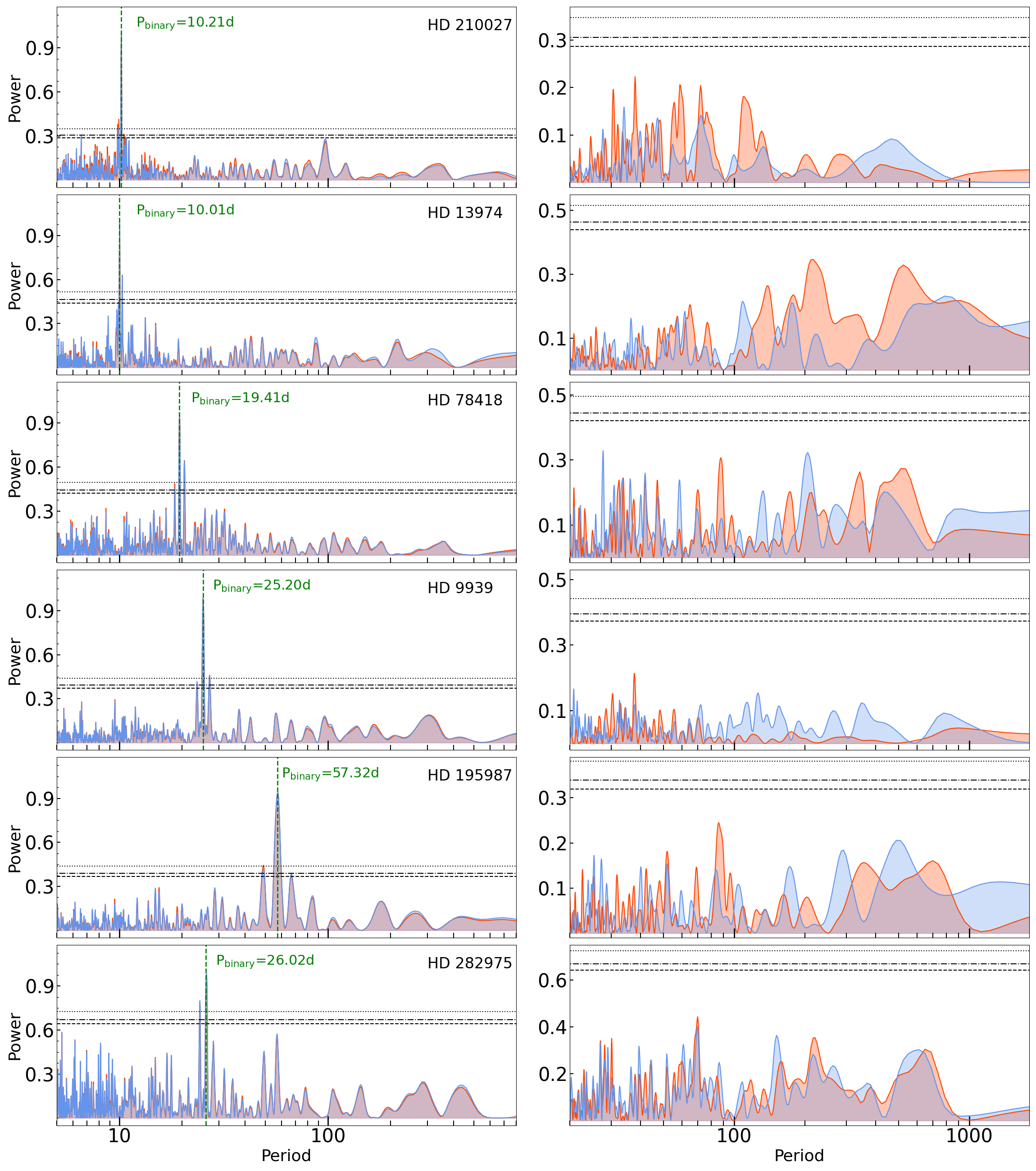}
\caption{Left panels: Periodogram of radial velocities for the primary (red) and the secondary stars (blue) with binary signal. Right panels: Periodograms after removing the binary motion.
The three horizontal lines indicate 10\% (dashed line), 5\% (dotted dash line), and 1\% (dotted line) false alarm probabilities.
}
\label{fig:period_all}
\end{figure*}

\section{A summary of the {\tt DOLBY} method of radial-velocity extraction}\label{sec:method}

In \cite{lalitha_2023}, we developed two data-driven approaches for accurately measuring radial velocities of double-lined binary systems (SB2s), that we now name {\tt DOLBY} (DOuble-Lined BinarY). Both methods make use of Gaussian Processes (GP) to disentangle both stellar components. The first method works in wavelength space, and makes a spectral decomposition ({\tt DOLBY}-SD), by modelling the observed spectra of a double-lined binary star as a sum of two GPsDoppler-shifted by their respective radial velocites (the quantity of interest). The second method works in velocity space, using the CCF produced by the SOPHIE pipeline ({\tt DOLBY}-CCF). Here both components are each modelled as the sum of GPs and of a Gaussian function. The {\tt DOLBY}-CCF approach is well suited for instruments such as HARPS, SOPHIE and ESPRESSO that produce well characterised and stable CCFs. 

In {\tt DOLBY}-SD, we divide the observed spectra into smaller wavelength subsets or "chunks" and applied GP regression separately to each chunk. This approach allows to account for different spectral types and optimises the reconstruction of the observed spectrum. The method uses a Mat\'ern kernel to model the covariance between the pixels. By exploring the posterior distribution of radial velocities of each chunk and the hyperparameters of the GP functions using a Markov Chain Monte Carlo sampler (MCMC), {\tt DOLBY}-SD computes refined estimates of the radial velocities for each component.

\begin{figure*}
\includegraphics[width=\textwidth]{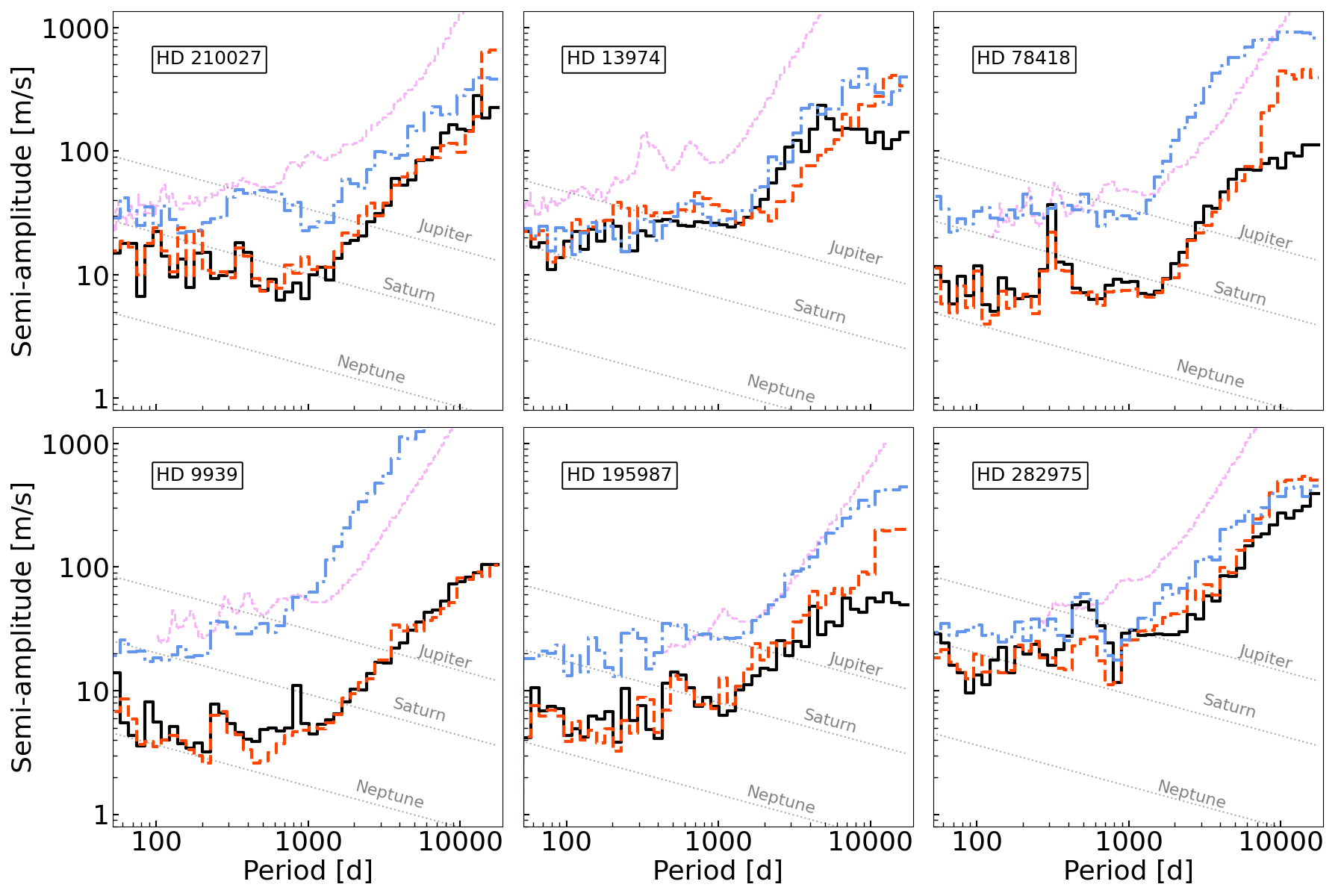}
\caption{The sensitivity limits based on the density of posterior samples for each binary system from the {\tt {kima}} run with N$_{\mathrm{P}}$ fixed to 1. Histograms lines show the amplitude at the 99th percentile within each period bin, giving us a 99\% confidence upper limit. The red and blue lines represent the analyses using the data solely on the primary and secondary stars respectively, while the black line represents the combined dataset. The violet curves indicate the planet detection limit from \citealt{Konacki_2009,Konacki2010}. Diagonal grey lines are anticipated signals of Neptune, Saturn and Jupiter mass planets orbiting stars with their respective binary masses.}
\label{fig:posteriors_all}
\end{figure*}

In {\tt DOLBY}-CCF, we instead employ a baseline mean function consisting of two Gaussian functions to capture each of the primary and secondary components of the double-lined binary star. In this method, the GPs are used to model the correlated wiggle signal of the CCF for each component using a Mat\'ern covariance kernel. The GP models are optimised using L-BFGS-B method \citep{Byrd1995} and the posterior distribution of hyperparameters are explored using MCMC sampling in the same way as is done for {\tt DOLBY}-SD.

Both these methods are effective in accurately measuring the radial velocities of double-lined binary systems. For more details on these methods and their implementation can be found in \cite{lalitha_2023}. For this paper we chose to use primarily the  {\tt DOLBY}-CCF method, which is by far the easiest to implement and more computationally efficient. This makes it the most likely to be used by other researchers. In the appendix, we provide Tables \ref{tab:HD210027_rvs} through \ref{tab:HD282975_rvs}, which contain the radial velocities obtained using {\tt DOLBY}-CCF method for the primary and secondary components of each of our targets.

\section{Results from fitting the radial-velocities}\label{sec:results}

The radial velocities produced by {\tt DOLBY}-CCF are fit using the \textit{BINARIES} model of {\tt kima} \citep{faria_kima_2018,baycroft_2023}. {\tt kima} is an orbital fitting algorithm utilising the diffusive nested sampler DNEST4 \citep{brewer_dnest4_2018}. {\tt kima} allows the number of Keplerian signals ($N_{\rm p}$) applied to the data to be a free parameter. As such we can perform model comparison by computing a Bayes' factor (BF) comparing a system with $N_{\rm p} = 0$ (a simple binary system) to $N_{\rm p} = 1$ (or higher numbers of planets too) while keeping all binary and planetary orbital parameters free. {\tt kima} fits for a systemic velocity, and also a radial-velocity jitter term which is added in quadrature to the uncertainties measured from {\tt DOLBY}-CCF, and is used to account for scatter produced by any unmodelled effects such as stellar variability. The \textit{BINARIES} model takes a different prior for the binary parameters as the planetary ones, using {\tt kima}'s known-object mode. Apsidal precession of the binary orbit is included as an additional free parameter, and a correction to the radial-velocities accounting for relativistic effects\footnote{These being the light-travel time, transverse doppler, and gravitational redshift effects \citep{zucker_spectroscopic_2007,sybilski_non-keplerian_2013}} is included \citep{baycroft_2023}. The \textit{BINARIES} model fits SB1 and SB2 radial velocity data equally. In the SB2 case (used in this work), {\tt kima} adjusts models to both the primary and secondary radial velocity time-series simultaneously. A separate systemic velocity and jitter are modelled for each of the primary star and secondary star. We use a Student's t distribution in the likelihood evaluation to account for any outliers in the radial velocities. In Fig. \ref{fig:phaseres} and \ref{fig:restphaseres}, we present the radial velocity time-series for each target, along with the best-fitting binary Keplerian models.

After analysing all six double-lined binaries, we do not detect any planets that pass our detection threshold, BF $\geq 150$ for a 1-planet model over a 0-planet model \citep{trotta_bayes_2008}. We thus consider only the posterior samples where no planet is included to derive the orbital parameters of each of the binaries. We report the fit parameters and Bayes' factors in Table \ref{tab:tab1}. A final advantage of a nested sampler is that it can be used to compute accurate sensitivity limits. We describe that procedure in section \S\ref{sec:limit}.

Being conscious that nested samplers remain a fairly recent occurrence in the exoplanet scientific literature, we also produce a more classical test. We compute a series of generalised Lomb-Scargle (GLS) periodograms \citep{Zechmeister_2009}. One periodogram is created for the radial velocities measured of each of the primaries, and for each of the secondary stars (see Figure \ref{fig:period_all} left panels). Any orbiting object should produce a similar signature in both the primary and secondary sets of radial-velocities, whereas parasitic signals such as stellar variability are expected to only show in one of the two components. The primary and the secondary stars are depicted in red and blue colours, respectively. We use a 10\,000  bootstrap randomisation of each input datasets to compute False Alarm Probabilities (FAP) levels of 10\%, 5\% and 1\%. The significant peaks shown in the left hand side columns of periodograms are associated with the binary star. 

To test for the presence of circumbinary exoplanets we compute a second series of GLS, after having removed the binary's orbital motion. To do that we subtract the best fitting Keplerian motion from the {\tt kima} analysis for each of the binary stars (Figure \ref{fig:period_all} right panels). There are no significant peaks in any of the radial-velocity residuals (right-hand column) that are consistent between the primary and secondary, and also consistent with the results obtained with {\tt kima}.

We now describe the results for every system individually.

\subsection{HD~210027}
HD~210027 is a bright F5V/G8V spectral type binary system with an orbital period of 10.2 d \citep{Konacki2010}. The SOPHIE observation period spans $\sim$800 days. The periodograms for HD~210027 in Fig.\ref{fig:period_all} (left panel) shows excess at binary period at $P_{\rm bin}\sim10.21~\rm d$ for both the primary and the secondary star with a ${\rm FAP}$ value significantly lower than $< 10^{-6}$. Subtracting this signal from the data and recalculating the periodogram (right panel) reveals a few moderate peaks at $5\times$ and $8\times$ the binary period for the primary star's radial velocities. These peaks have a False Alarm Probability (FAP) of less than 0.05. %However, no significant peaks are found in the secondary star's radial velocities. 
The Bayes' Factor for the $N_{\rm p} = 1$ model vs $N_{\rm p} = 0$ model is $\sim$1.6. We therefore do not associate this signal with a planetary signal, but possibly with stellar activity on the primary star instead. %\cite{Konacki_2009} characterised the system by a root mean square (rms) of $17.1 \rm ~m\,s^{-1}$ for the primary and $69.7 \rm ~m\,s^{-1}$ for the secondary. 
Our radial velocity solution has rms of $12.5 \rm ~m\,s^{-1}$  and $40.3 \rm ~m\,s^{-1}$ for the primary and secondary stars, respectively. The $M \sin^3 i$  for the primary and secondary are $1.31174\pm0.00031 \, \rm M_\odot$ and $0.81769\pm0.00013\, \rm M_\odot$, with a precision of 
0.027$\%$ and 0.015$\%$, respectively. Compared to previous work, the mass estimates are $\sim59\%$ and $\sim77\%$ more precise for the primary and secondary stars.

\subsection{HD~13974}
HD~13974 is a 10-d period binary with a spectral type of G0V \cite{Konacki_2009}. With an observation span extends over $\sim$ 1100-d, the periodograms for HD 13974 (Fig.\ref{fig:period_all}, left panel) reveal an evident excess at the binary period of approximately $P_{\rm bin}\sim10.0177$ days for both the primary and secondary star with FAP $< 10^{-6}$. After removing the strong binary signal from the data, the periodogram is rerun on this binary-subtracted data (shown in the right panel of Fig.\ref{fig:period_all}). This reveals that the primary star's radial velocity data still exhibits a peak, but the associated FAP is now less than $0.05 (5\%)$. %However, there are no notable peaks detected in the periodogram of the radial velocities for the secondary star.  
The BF for the $N_{\rm p} = 1$ model vs $N_{\rm p} = 0$ model is $\sim$6.6. Our radial velocity solution has an rms of 18.1 $\rm\,m\,s^{-1}$ for the primary star and 24.9 $\rm\,m\,s^{-1}$ for the secondary star. %, showcasing improvements over the previous work by \citep{Konacki_2009}, which reported rms values of 22.5 $\rm,m,s^{-1}$ and 111.1 $\rm,m,s^{-1}$, respectively.

\subsection{HD~78418}
HD 78418 is a G5 spectral type binary system with 19.4 d orbital period \citep{Konacki2010}. 
The periodograms for HD~13974 in Figure \ref{fig:period_all} (left panel) reveal a notable signal at a binary period of approximately $19.3971$ d for both the primary star and with an associated FAP $< 10^{-6}$. When we subtract this signal and re-run the periodogram, as shown in the right panel, we observe a peak at 87.88 d  for the primary star and $\sim$205 d with FAP $<0.05 (15\%)$ for the secondary star. However, due to the absence of significantly prominent peaks in both stars, we refrain from making definitive comments on the significance of these signals at this stage. The BF for the $N_{\rm p} = 1$ model vs $N_{\rm p} = 0$ model is $\sim$2.1. Nevertheless, we intend to continue monitoring this system. The duration of SOPHIE observations spans over 1135 days. The radial velocities derived by our method have an rms of 6.2 and 21 $\rm,m,s^{-1}$ for the primary and the secondary stars, respectively.%,  compared to 11.5 $\rm,m,s^{-1}$ and 25$\rm,m,s^{-1}$ m/s reported by \citep{Konacki_2009}.

\subsection{HD~9939}
HD 9939 comprises a binary system characterised by a K0IV spectral type, with an orbital period of 25.2 d \citep{Boden2006}. The SOPHIE observation period spans over 1138 d. 
The periodograms for HD~9939 in Fig.\ref{fig:period_all} (left panel) shows excess at binary period at $\sim$25.2319 d for both the primary and the secondary stars with a FAP $< 10^{-6}$. The radial velocity determined using our methodology, exhibits rms  of 5.8 $\rm\,m\,s^{-1}$ for the primary star and 63.4 $\rm\,m\,s^{-1}$ for the secondary star. %In a related study, \cite{Konacki_2009} characterised HD~9939 with rms values of 19.5 $\rm\,m\,s^{-1}$ for the primary star and 36 $\rm\,m\,s^{-1}$ for the secondary star. 
%Note that our method outperforms the characterisation of the primary star by approximately $\sim$70\%. Interestingly, for the secondary star, our method appears to be less effective. Here, it is relevant to highlight that this particular 
We note that the scatter in radial velocity measurements is the largest among all the targets, and we speculate it might be caused by intrinsic variability of this particular star. 

The BF for the $N_{\rm p} = 1$ model vs $N_{\rm p} = 0$ model is $\sim$0.4. When we subtract the Keplerian signal with a period of $\sim$25.2319 days, as depicted in the periodogram (right panel), we observe two peaks at around 38 and 48 days with a FAP exceeding $\sim0.05 (5\%)$ in the radial velocities of the primary star. Stellar variability could potentially be the source of this signal, which warrants further investigation. %However, a relatively weaker signal is detected in both the primary and secondary stars at $\sim$121 days. The reduced significance of this signal in the residual radial velocity of the secondary star might be attributed to the substantial scatter that is discussed further up. Further observations of this target may provide more information regarding the origin and significance of this signal.

\subsection{HD~195987}
HD 195987 is a G3V/K2V binary system with an orbital period of 57 days \citep{Torres2002}, making it the system with the longest binary period within our set of targets. Our observational coverage with SOPHIE spans $\sim$766 days. In the periodogram analysis, a signal emerges at the binary period of $\sim57.3129$~d for both the primary and secondary stars with a FAP $< 10^{-6}$. Upon removing this signal from the data, we do not detect any other prominent peaks in the periodogram (Fig. \ref{fig:period_all}, right panel). The BF for the $N_{\rm p} = 1$ model vs $N_{\rm p} = 0$ model is $\sim$1.2. In the case of HD 195987, our radial velocity solution produces a rms of 4.8 $\rm\,m\,s^{-1}$ for the primary star, effectively reaching the photon noise level, and 27.3 $\rm\,m\,s^{-1}$ for the secondary star.
%,  compared to 11 $\rm,m,s^{-1}$ and 48 $\rm,m,s^{-1}$ reported by \citep{Konacki_2009}.

\subsection{HD~282975}
HD 282975 is a binary system featuring G6V stars with an orbital period of 26 days \citep{Konacki_2009}. 
The duration of our survey extends to 1107 days. The periodogram analysis of the radial velocity data from HD 282975 produces signal at a binary period of $\sim$26.1681 days for both the primary and the secondary stars with  FAP $< 10^{-6}$. Following the removal of this prominent peak, no other significant peak emerges (Fig. \ref{fig:period_all}, right panel.
%a moderately significant signal emerges at approximately 269 days with a 10\% FAP for the primary star. However, no similarly significant peak is detected for the secondary star. 
The BF for the $N_{\rm p} = 1$ model vs $N_{\rm p} = 0$ model is $\sim$40.6, a value below the detection threshold of 150. Furthermore, the results of our radial velocity analysis produces an rms of 10.6 $\rm\,m\,s^{-1}$ for the primary star and 20.6 $\rm\,m\,s^{-1}$ for the secondary star.

\begin{figure*}
\includegraphics[width=\textwidth]{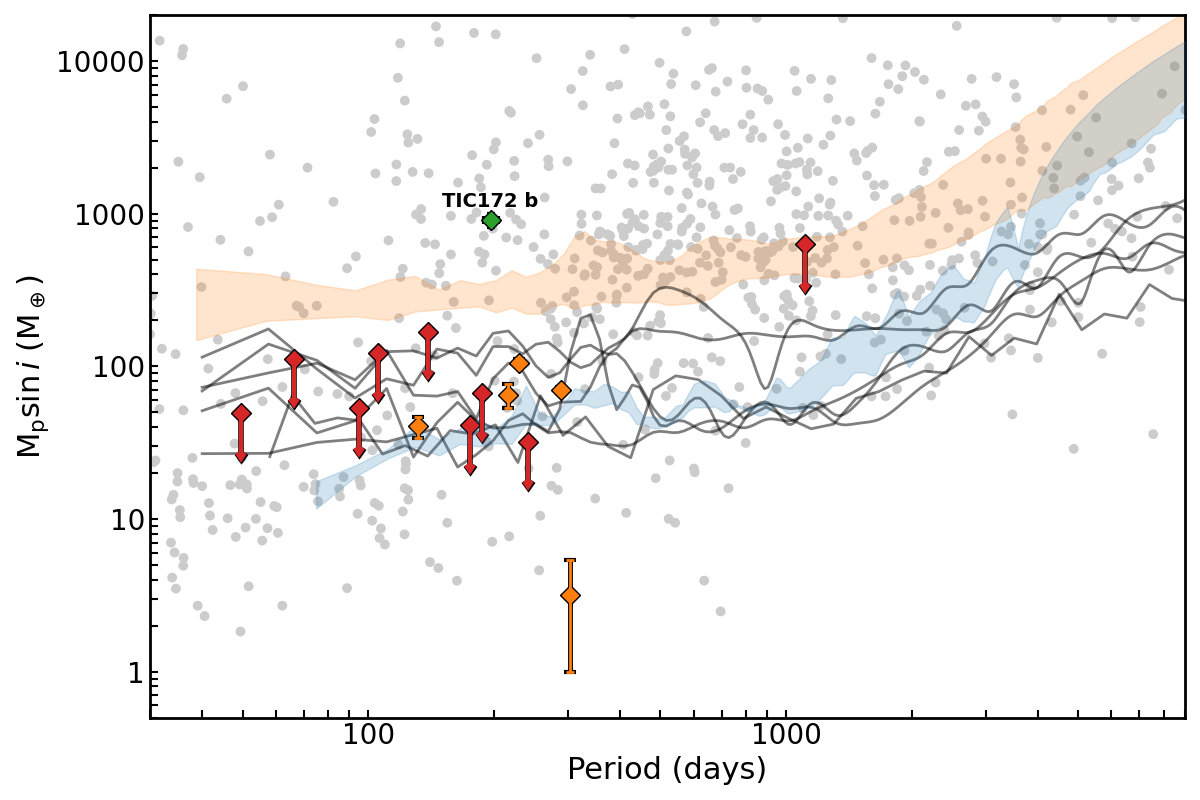}
\caption{Minimum planet mass versus period plot showcasing the detection limits of {\tt DOLBY} targets in this study, represented as black lines, alongside confirmed exoplanets as grey circles, and circumbinary planets as orange/red diamonds. Circumbinary planets with upper mass limits are marked with arrows signifying a 2$\sigma$ upper limit. The blue band denotes the detection limits for SB1 systems \citep{standing_2022}, while the orange band represents the detection limits for SB2 systems \citep{Konacki_2009,Konacki2010}. TIC 172900988b is depicted as a green diamond \citep{lalitha_2023}.}
\label{fig:detection_limit_discussion}
\end{figure*}

\subsection{Detection limit}\label{sec:limit}
The use of {\tt kima} (and nested sampling in general) lends itself to calculating detection limits in a single analysis. In cases where no planet is detected, as is the case for the systems in this work, we fix the number of Keplerian signals included in the model to one and perform another analysis. The posterior samples resulting from this analysis will then correspond to planetary signals that are consistent with the data but not formally detected. To obtain a detection limit the posterior samples are binned in period with bins of equal width in log-space, then within each bin the detection limit is the value of the semi-amplitude $K$ below which $99\%$ of the posterior samples lie. A more detailed explanation of this method can be found in \citet{standing_2022}. To ensure that the detection limit is robust we ensure that each period bin contains at least 1000 posterior samples.

For this work, we perform three detection limits for each binary, the first taking only the radial velocity data for the primary and fitting as a single-lined binary, the second taking only the data for the secondary, and the third taking both and fitting as a double-lined binary. These are shown in Figure \ref{fig:posteriors_all} alongside the detection limits from \citet{Konacki_2009}. The diagonal lines indicate the expected signal amplitudes of Neptune, Saturn, and Jupiter mass planets orbiting each binary. We see that our detection limits are consistently more sensitive, up to an order of magnitude better, in the best case. This could come from DOLBY being an improved method at extracting radial-velocities, from an increase in data between \citet{Konacki2010} and our study, and/or from a difference in how the detection limits are estimated.

To further evaluate the performance of our radial-velocity extraction method, we check the influence of data amount and computation of the detection limit. To test the detection limit methods, we first focus on two specific systems, HD 210027 and HD 78418. These systems were chosen because they are the only ones for which we have available radial velocity data from \citet{Konacki2010}. We have applied the {\tt kima} method to the radial velocity data from \citet{Konacki2010} for these systems to obtain detection limits. These results are compared with the detection limits derived from our {\tt kima-DOLBY} analysis of the same data and with the original detection limits reported by \citet{Konacki_2009}. Figure \ref{fig:kima_compare} shows the comparison of these detection limits for both systems. The solid black lines represent the detection limits obtained from our {\tt kima-DOLBY} analysis, while the green lines correspond to the detection limits derived from the Konacki data using {\tt kima}. The pink lines indicate the original detection limits reported by \cite{Konacki_2009}. 

In our comparison of detection limits, we consistently observe that {\tt DOLBY} performs better than previous approaches. Given that we have now employed {\tt kima} on Konacki's radial velocity data and achieved consistent results, the improvement in detection limits is not attributable to the detection limit calculation method itself. Instead, the enhancement could likely be from either the superior accuracy of the {\tt DOLBY} method or the increased number of measurements.

To quantify the impact of the number of observations, we systematically scaled the Konacki detection limits based on the square root of the relative number of measurements for each system, shown in Figure~\ref{fig:detection_limit_scaled}. This scaling assumes that the detection limit improves proportionally with the square root of the number of observations. Our analysis reveals that even after accounting for this scaling, the {\tt DOLBY} method still demonstrates an improved performance, suggesting that its inherent accuracy contributes significantly to the observed improvement in detection limits.

\subsection{Comparison between methods and datasets}

Across the six systems analysed, the mean radial-velocity error for each system ranges between 4.2-11.6 ${\rm m\,s^{-1}}$ for the primaries and 9.3-44.3 ${\rm m\,s^{-1}}$ for the secondaries. In addition, we measure an RMS scatter after removing the most likely binary solution ranging from 4.8-18.1 and 20.6-63.4 ${\rm m\,s^{-1}}$ for primaries and secondaries respectively. In all but 3 of the 12 cases the radial velocity jitter that is included in the fit is not constrained and is consistent with 0. This shows that we are reaching a regime where we are photon-noise limited in the best cases. It is also worth noting here that all of the 6 binaries have at least one of the primary and secondary radial velocities close to photon-noise, reinforcing the notion that observing double-lined binaries can be a favourable way to avoid stellar activity relative to single stars or single-lined binaries. When only one component is resolved, and if that component is active, then detecting exoplanets becomes challenging. In the case of double-lined systems, the activity cycle of both stars does not need to be related, and while one star is active, the other star might not be, providing a way to continue the radial-velocity monitoring.

% Across the six systems we analysed, we obtained radial-velocity error weighted precision from 6 to 37 $\rm\,m\,s^{-1}$ for the primary stars, and 7 to 58 $\rm\,m\,s^{-1}$ for the secondary stars.
% In addition, we measure an RMS scatter after removing the most likely binary solution that ranges from 4.8 to 18.1 $\rm\,m\,s^{-1}$ for primary stars, and 20.6 to 63.4 $\rm\,m\,s^{-1}$ for secondary stars. In only two out of the twelve cases, we find consistency between the root mean square (RMS) and the radial-velocity precision measured with {\tt DOLBY}. This consistency is observed for the secondary stars of HD 78418 and HD 195987, where we achieve 23 $\rm\,m\,s^{-1}$ and 27 $\rm\,m\,s^{-1}$, respectively. These measurements were obtained over timeseries comprising 37 and 52 radial-velocity measurements, respectively.
% This implies that either the {\tt DOLBY}-CCF uncertainties are under-estimated, or that double-lined binary systems intrinsically produce a radial-velocity scatter. Further development of the {\tt DOLBY} method is planned, and it is expected that {\tt DOLBY}-SD which was not used here, would produce more accurate and precise results than {\tt DOLBY}-CCF.

The uncertainties and jitters are shown in Figure \ref{fig:uncertainity}, where the box plot showcases the distribution of uncertainties for each binary system and compares them to other results.
%\footnote{Note: When specifically considering the HIRES data, we denote it as 'Konacki (HIRES)'. When considering all combinations of telescopes (Keck/Hires, SHANE/CAT/Hamspec, and TNG/SARG), we denote it as '{\tt kima}-Konacki'.}

We note that {\tt DOLBY}-CCF systematically outperforms TODMOR \citep{Zucker_2004}, the only public code we could directly compare our results to, and analyse the same data with. In addition, the SOPHIE sample analysed with {\tt DOLBY} produces uncertainties that are systematically smaller than results reported in \citet{Konacki_2009,Konacki2010} for the same systems. Also, the {\tt DOLBY} results typically cover a narrower range of values (i.e, measurements are more consistent with one another). RMS values of the RV residuals are also depicted. %In seven out of ten cases, the RMS obtained with SOPHIE+{\tt DOLBY} is smaller than those reported in \citet{Konacki_2009,Konacki2010} despite using a 2m telescope as opposed to a 10m (Keck), and while collecting on average twice as many measurements for each system. 
Despite using a 2m telescope and collecting on average twice as many measurements for each system, the detection sensitivty obtained with SOPHIE+{\tt DOLBY} is better in all cases than those reported in \citet{Konacki_2009,Konacki2010}, even though Konacki et al. used a combination of telescopes including the 10m Keck/Hires, SHANE/CAT/Hamspec, and TNG/SARG.
We take this as evidence that the {\tt DOLBY} method is slightly more accurate than the disentangling method of \citet{Konacki_2009, Konacki2010}. %but do note that a change in stellar activity between the TATOOINE observations \citep[which are 15 years old;][]{Konacki_2009,Konacki2010}, and the SOPHIE observations could also have caused the discrepancy.

\section{Conclusions}\label{sec:conclusion}

This paper presents the application of a novel method, called  {\tt DOLBY}-CCF, to measure radial-velocities of double-lined binary stars with high precision. We applied {\tt DOLBY}-CCF to analyse newly collected SOPHIE observations of six double-lined systems previously observed by the TATOOINE survey \citep{Konacki_2009,Konacki2010}. Our goal was to assess the capability of {\tt DOLBY}-CCF for accurate RV measurements of double-lined binary stars stars, and to assess its potential to detect circumbinary exoplanets.

We achieved a significant improvement in RV precision compared to the traditional TODMOR method (Fig.\ref{fig:uncertainity}). 
Our analysis did not reveal any significant detections of circumbinary planets within our sample. Figure \ref{fig:detection_limit_discussion} depicts a plot of mass versus the orbital period of detection limits obtained from this work for double-lined binaries (black). It also includes detection limits for single-lined binaries (SB1; blue band) obtained by \citet{standing_2022} and confirmed exoplanets (grey filled circles; NASA Exoplanet Archive\footnote{\href{https://exoplanetarchive.ipac.caltech.edu/index.html}{https://exoplanetarchive.ipac.caltech.edu/index.html}}). Known transiting circumbinary planets with mass measurements are represented as filled diamonds. We also show the range of detection limits reported by \citet{Konacki_2009,Konacki2010} as an orange band. 
While detection limits for single-lined binaries by \citet{standing_2022} are compatible to the {\tt DOLBY}-CCF method 
for SB2s, our analysis offers up to one order-of-magnitude improvement in sensitivity compared to the previous results reported in \citet{Konacki_2009,Konacki2010}. This improvement could come from various sources: 1) greater accuracy in the radial-velocity extraction, 2) more radial-velocity measurements, and 3) improved methods to compute sensitivity limits, however, by analysing publicly available data from \citet{Konacki2010}, and rescaling detection limits to account of differences in the number of measurements, we show that the main improvement is most likely because {\tt DOLBY}-CCF extracts more accurate radial-velocities than previous methods. Figure \ref{fig:detection_limit_discussion} also showcases it is now possible to detect circumbinary planets with masses below that of Saturn for orbital periods of up to 1000 days—mirroring the detection capacity observed in single-lined binaries \citep{standing_2022}. 
The high RV precision achieved by {\tt DOLBY}-CCF paves the way for future, more sensitive 
searches for circumbinary planets around double-lined binary stars. 

Future work will involve applying {\tt DOLBY}-CCF to a larger sample of double-lined binary stars and incorporating additional observations to increase the sensitivity of our planet searches. In addition, we aim to test the {\tt DOLBY}-SD method to compare it to the CCF approach.
Our current results however demonstrate  that {\tt DOLBY} represents a valuable, public tool capable of advancing our understanding of circumbinary systems and their potential to harbour exoplanets.

\section*{Acknowledgements}

This paper is based on observations collected at the Observatoire de Haute Provence (OHP). We are grateful to the entire staff at OHP, especially the night assistants, for their dedication and hard work in securing these observations, particularly during the challenging times of the COVID pandemic.

This research was funded by the European Research Council (ERC) under the European Union's Horizon 2020 research and innovation programme (grant agreement no. 803193/BEBOP) and by the Leverhulme Trust (research project grant no. RPG-2018-418). PM acknowledges support from STFC research grant number ST/S001301/1. MRS
acknowledges support from the European Space Agency as an ESA Research Fellow. IB, GH, AS, MD ackowledge the Programme National de Planetologie (PNP)of the CNRS INSU co-funded by CNES  for the funds received to support this work. The acquisition of data was facilitated by a series of allocations through the French PNP. The computations for this research were performed using the University of Birmingham's BlueBEAR HPC service \href{http://www.birmingham.ac.uk/bear}{http://www.birmingham.ac.uk/bear}.

\afterpage{%
    \begin{landscape}
    \begin{table}
    \centering
    
    \caption{Properties of SB2s}
    \label{tab:tab1}
    \begin{tabular}{lllllll}
\hline\hline
    Parameters & HD~210027 & HD~13974 & HD~78418 & HD~9939 & HD~195987 & HD~282975 \\
    \hline\\

    V (mag) & 3.76 & 4.9 & 5.98 & 6.99 & 7.09 & 10.0 \\
    Spectral type & F5V/G8V & G0V & G5 & K0IV & G3V/K2V & G6V \\
    Effective temperatures (K) & 6642/4991 & & 6000/5900 & && \\
    %P$_{\mathrm{binary}}$ (days) & 10.2 & 10 & 19.4 & 25.2 & 57 & 26 \\
    Number of epochs & 56 & 35 & 37 & 43 & 52 & 21 \\
    $i$ & $95.83\pm 0.12 $ & --- & $146.88 \pm 0.25$ & $61.56 \pm 0.25$ & $99.364\pm0.080 $ & ---\\
    \hline
    Fit parameters &&&&&&\\
    $P$ & $10.2130447 \pm 0.0000040$ & $10.020195\pm0.000020$ & $19.412378^{+0.000010}_{-0.000014}$ & $25.208811^{+0.000015}_{-0.000018}$ & $57.321927^{+0.000093}_{-0.000065}$ & $26.04626\pm0.00035$ \\
    $e$ & $0.00183846^{+0.000043}_{-0.000046}$ & $0.0093\pm0.00047$ & $0.194966^{+0.000050}_{-0.000054}$ & $0.101283^{+0.000047}_{-0.000045}$ & $0.305100^{+0.000034}_{-0.00031}$ & $0.25051^{+0.00048}_{-0.00056}$ \\
    $w$ & $4.730^{+0.024}_{-0.015}$ & $6.188^{+0.042}_{-0.033}$ & $4.94798^{+0.00045}_{-0.00042}$ & $5.45092^{+0.00068}_{-0.00061}$ & $6.23339^{+0.00011}_{-0.00018}$ & $1.5652^{+0.0054}_{-0.0060}$ \\
    $K$ & $48479.0^{+2.5}_{-2.2}$ & $10099.4\pm4.5$ & $26488.66^{+0.97}_{-1.30}$ & $35116.0\pm1.4$ & $28852.48^{+0.89}_{-0.99}$ & $8169.5^{+4.7}_{-4.3}$ \\
    $q$ & $0.623362^{+0.000083}_{-0.000065}$ & $0.72090^{+0.00045}_{-0.00050}$ & $0.86095^{+0.00017}_{-0.00013}$ & $0.78692^{+0.00014}_{-0.00011}$ & $0.786314^{+0.000119}_{-0.000098}$ & $0.94962^{+0.00068}_{-0.00075}$ \\
    $\dot{\omega}$ & $0\pm1000$ & $0\pm1000$ & $158\pm62$ & $-60\pm100$ & $21^{+34}_{-27}$ & $340\pm700$ \\
    ${\rm T}_{\rm per}$ & $2459483.099^{+0.039}_{-0.024}$ & $2459300.427^{+0.068}_{-0.051}$ & $2459409.0138\pm0.0014$ & $2459374.6961^{+0.0026}_{-0.0023}$ & $2459378.81855^{+0.00072}_{-0.00191}$ & $2459442.694\pm0.022$ \\
    ${\rm V}_{\rm sys, pri}$ & $-4800.8\pm2.1$ & $-6558.4^{+3.9}_{-4.3}$ & $9608.6^{+2.2}_{-2.0}$ & $-46047.8\pm1.2$ & $-5630.6\pm1.0$ & $6597.0\pm4.5$ \\
    ${\rm V}_{\rm sys, sec}$ & $-4240.2\pm6.1$ & $-6347.3\pm6.2$ & $9961.6^{+4.1}_{-3.9}$ & $-45670.7^{+6.0}_{-5.5}$ & $-5616.1\pm4.0$ & $7659.2^{+5.1}_{-4.9}$ \\
    \hline
    Derived parameters\\
    ${\rm M}_{\rm pri}\,\sin^3(i)$ & $1.31174\pm0.00031$ & $0.0084530\pm0.0000095$ & $0.191406\pm0.000067$ & $0.72977\pm0.00026$ & $0.80864\pm0.00024$ & $0.0059273^{+0.0000098}_{-0.0000102}$ \\
    ${\rm M}_{\rm sec}\,\sin^3(i)$ & $0.81769\pm0.00013$ & $0.0060937\pm0.0000060$ & $0.164793^{+0.000032}_{-0.000034}$ & $0.57427\pm0.00011$ & $0.63585\pm0.00011$ & $0.0056283\pm0.0000077$ \\
    
    ${\rm M}_{\rm pri}$& $1.33233\pm0.00092$ &  ---  & $1.173\pm0.024$ & $1.0733\pm0.0076$ & $0.84184\pm6.4{\rm e-}04$ & --- \\
    ${\rm M}_{\rm sec}$& $0.83051\pm5.5{\rm e-}04$ & --- & $1.010\pm0.021$ & $0.8446\pm0.0060$ & $0.66197\pm4.7{\rm e-}04$ & --- \\
    ${\rm a}_{\rm bin}$ & $0.119137\pm2.7{\rm e-}05$ & --- & $0.1834\pm0.0012$ & $0.20905\pm4.9{\rm e-}0.4$ & $0.333335\pm8.2{\rm e-}05$ & --- \\
    ${\rm a}_{\rm pri}$ & $0.045747\pm0.000010$ & --- & $0.08484^{+5.8{\rm e-}04}_{-5.6{\rm e-}04}$ & $0.09206\pm2.2{\rm e-}0.4$ & $0.146733\pm3.4{\rm e-}05$ & --- \\
    ${\rm a}_{\rm sec}$ & $0.073390\pm0.000018$ & --- & $0.09854^{+6.7{\rm e-}04}_{-6.5{\rm e-}04}$ & $0.11699\pm2.8{\rm e-}0.4$ & $0.186602\pm5.1{\rm e-}05$ & --- \\
    
    \hline
    Bayes Factor $N_{\rm p}=1/N_{\rm p}=0$  & 1.6 & 6.6 & 2.1 & 0.4 & 1.2 & 40.6\\
    mean $\sigma_{\rm RV, pri}$ & $11.6$ & $4.2$ & $4.3$ & $4.9$ & $4.4$ & $8.6$  \\
    mean $\sigma_{\rm RV, sec}$ & $44.3$ & $29.0$ & $14.8$ & $29.1$ & $9.3$ & $9.5$  \\
    rms$_{\rm pri}$ & $12.5$ & $18.1$ & $6.2$ & $5.8$ & $4.8$ & $10.6$ \\
    rms$_{\rm sec}$ & $40.3$ & $24.9$ & $21.0$ & $63.4$ & $27.3$ & $20.6$ \\
    Jitter$_{\rm pri}$ & $<4.0$ & $14.9\pm3.2$ & $<2.9$ & $<1.4$ & $<2.6$ & $<3.0$ \\
    Jitter$_{\rm sec}$ & $<8.3$ & $<5.1$ & $12.3^{+5.1}_{-9.3}$ & $<10.1$ & $25.6^{+3.4}_{-3.6}$ & $<12.6$ \\
    \hline
    \end{tabular}
    
\footnotesize{Note: Inclination values were obtained from previous literature sources: \cite{Boden2006} for HD 9939, \cite{Konacki_2009} for HD78418 and HD 210027, and \cite{Torres2002} for HD195987. No inclination measurements were found for HD 13974 and HD 282975.}
    \end{table}
    \end{landscape}
    }

%%%%%%%%%%%%%%%%%%%%%%%%%%%%%%%%%%%%%%%%%%%%%%%%%%
\section*{Data Availability}

The full radial velocity data for this study are provided in the appendix. Reduced spectra used in the analysis are publicly available through the SOPHIE archive \href{http://atlas.obs-hp.fr/sophie/}{http://atlas.obs-hp.fr/sophie/}.



\bibliographystyle{mnras}
\bibliography{paper}

%%%%%%%%%%%%%%%%%%%%%%%%%%%%%%%%%%%%%%%%%%%%%%%%%%

%%%%%%%%%%%%%%%%% APPENDICES %%%%%%%%%%%%%%%%%%%%%

\appendix

\section{Observed and modelled radial velocities}

\begin{figure}
\includegraphics[width=0.475\textwidth]{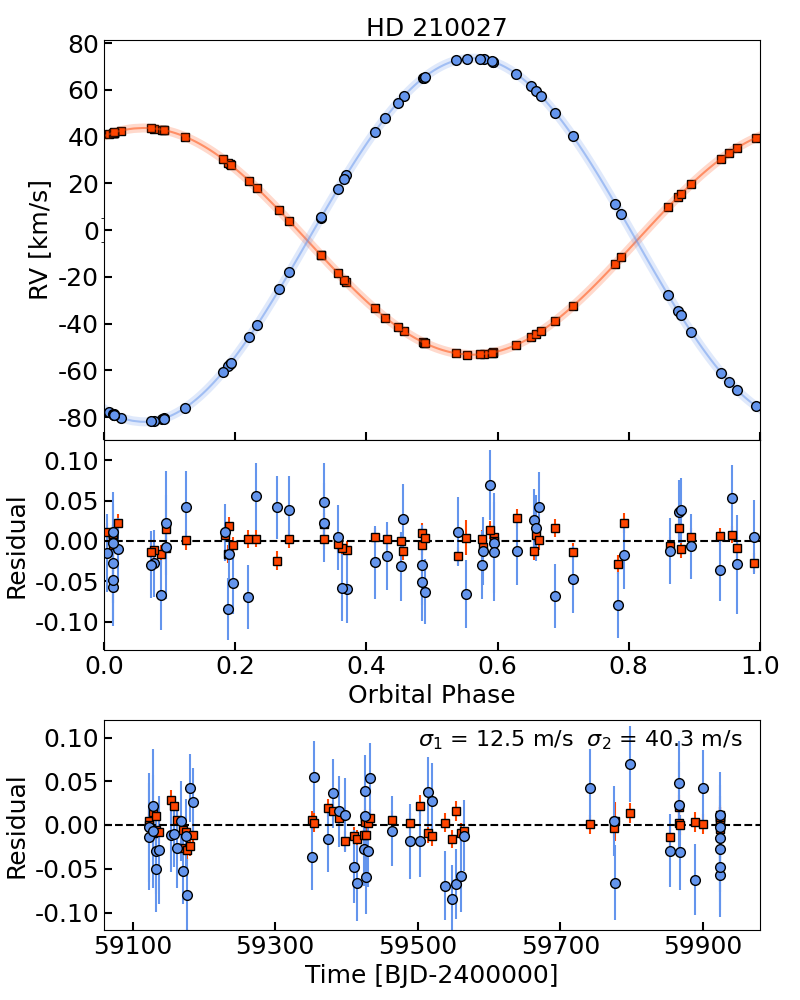}
\includegraphics[width=0.475\textwidth]{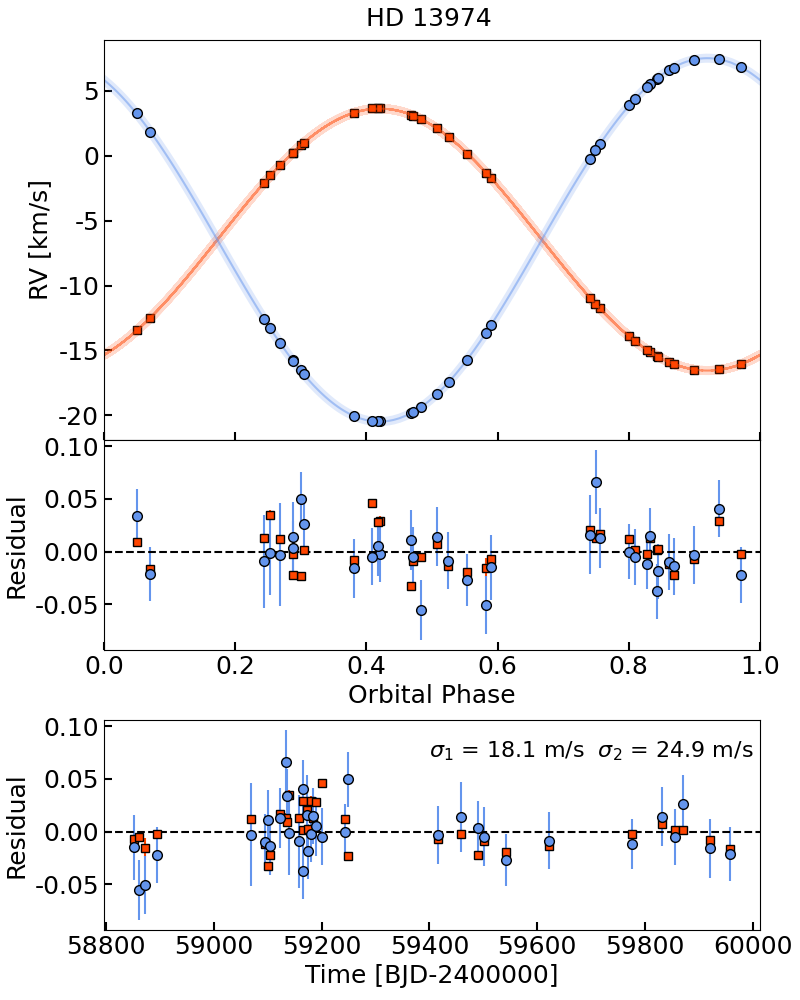}
\caption{ Observed and modelled radial velocities of HD 210027 and HD 13974 as a function of the orbital phase, their best-fit residuals as a function of the orbital phase  and time. The orange squares and the blur circle symbols represents the primary and the secondary stars, respectively.}
\label{fig:phaseres}
\end{figure}

\begin{figure*}
\includegraphics[width=0.475\textwidth]{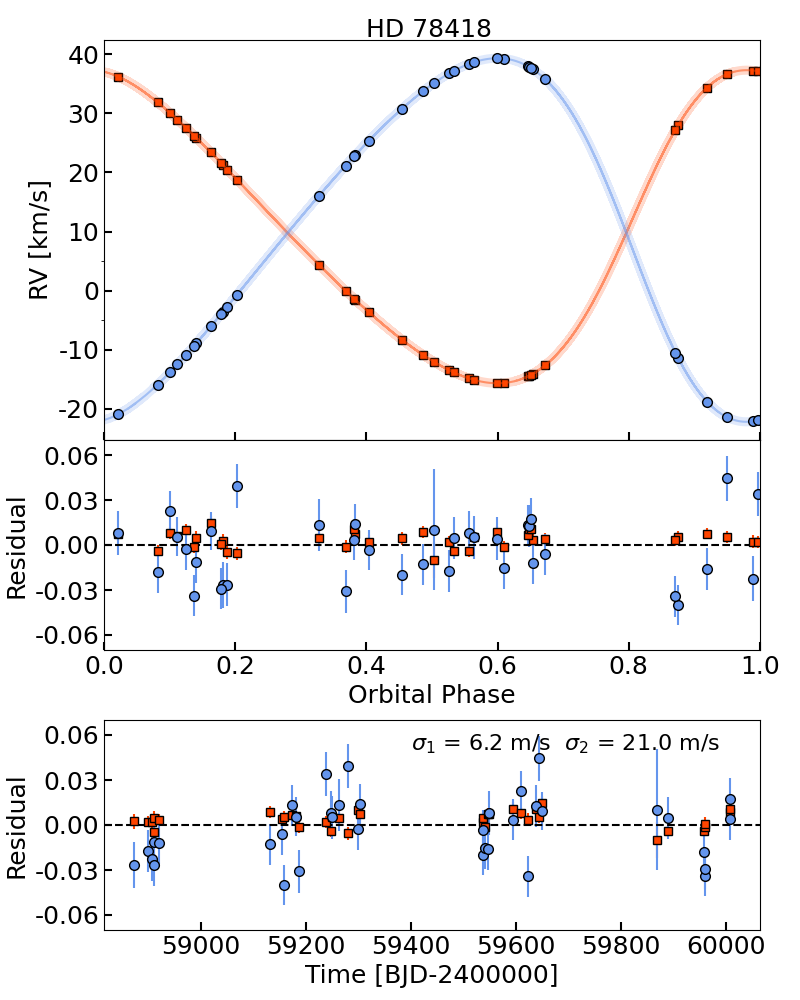}
\includegraphics[width=0.475\textwidth]{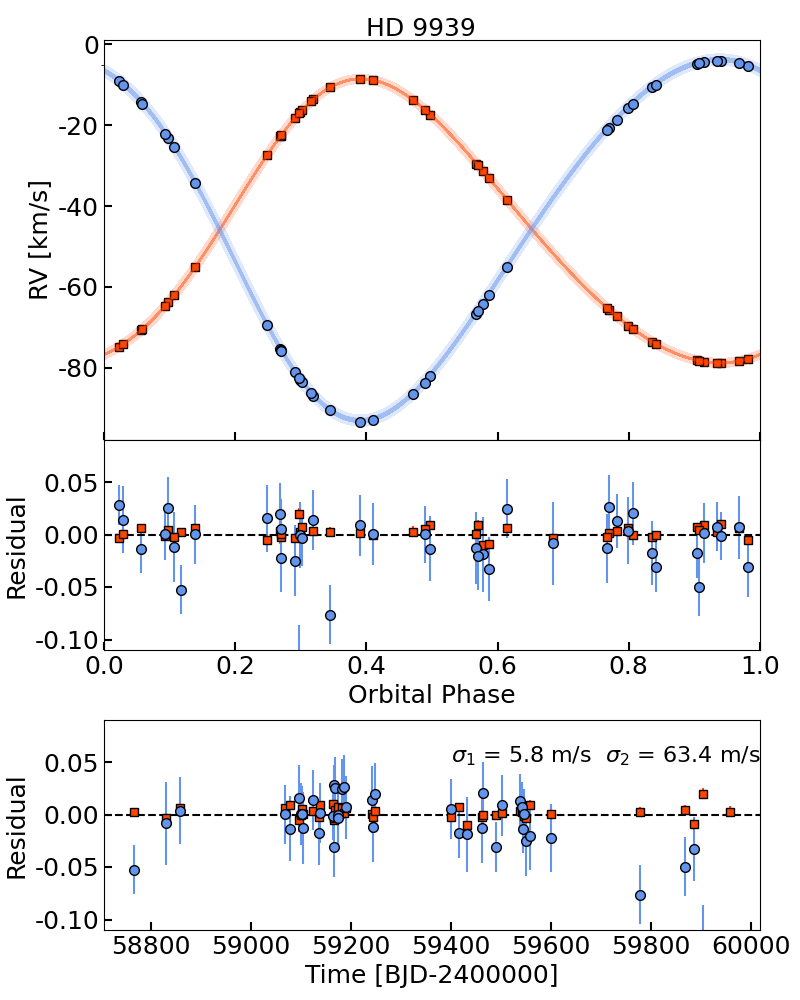}
\includegraphics[width=0.475\textwidth]{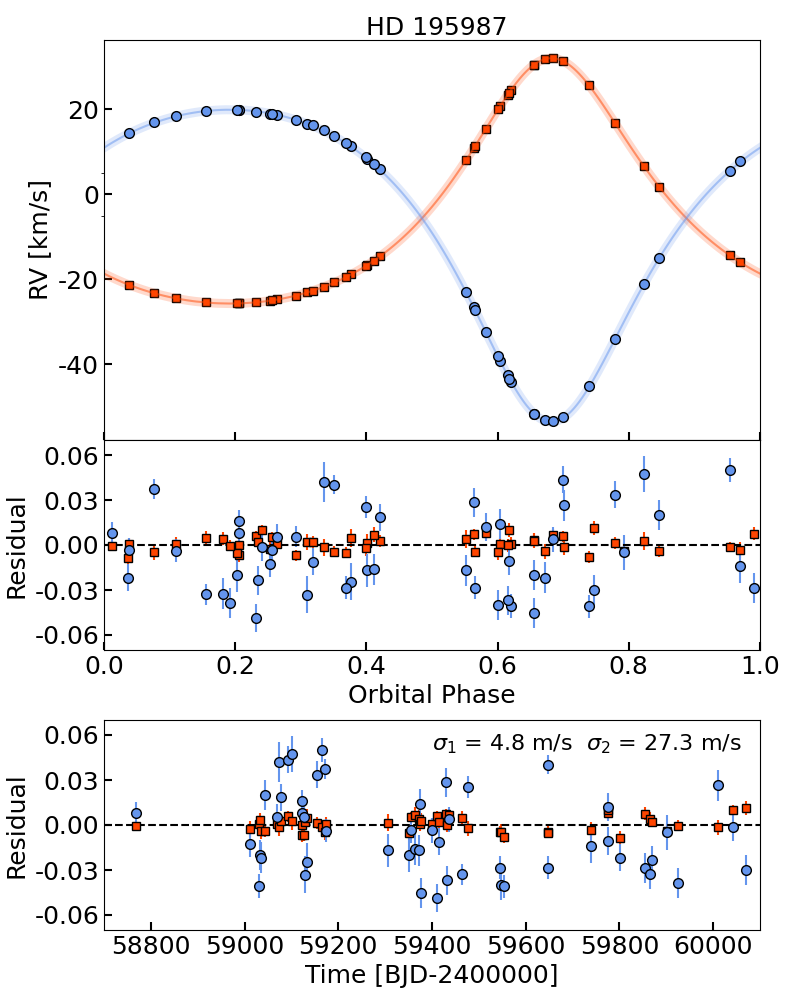}
\includegraphics[width=0.475\textwidth]{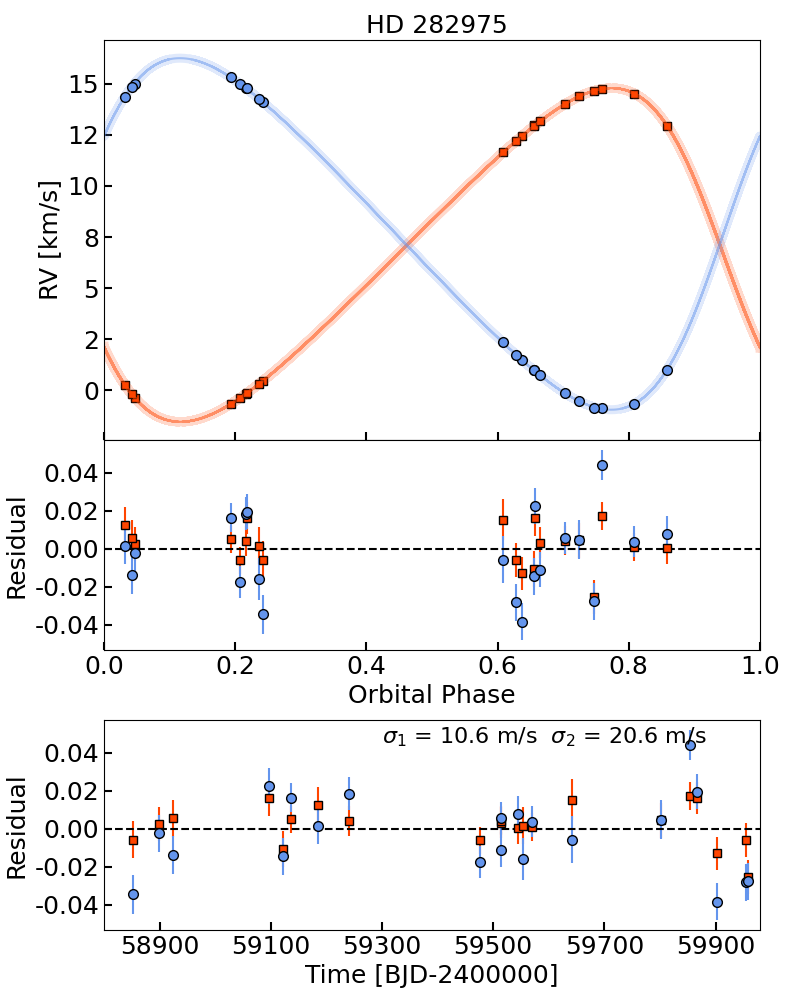}
\caption{Observed and modelled radial velocities of HD 78418, HD 9939, HD 195987 and HD 282975 as a function of the orbital phase, their best-fit residuals as a function of the orbital phase  and time. The orange squares and the blur circle symbols represents the primary and the secondary stars, respectively.}
\label{fig:restphaseres}
\end{figure*}

\section{Uncertainties and RMS values}

\begin{figure*}
\includegraphics[width=\textwidth]{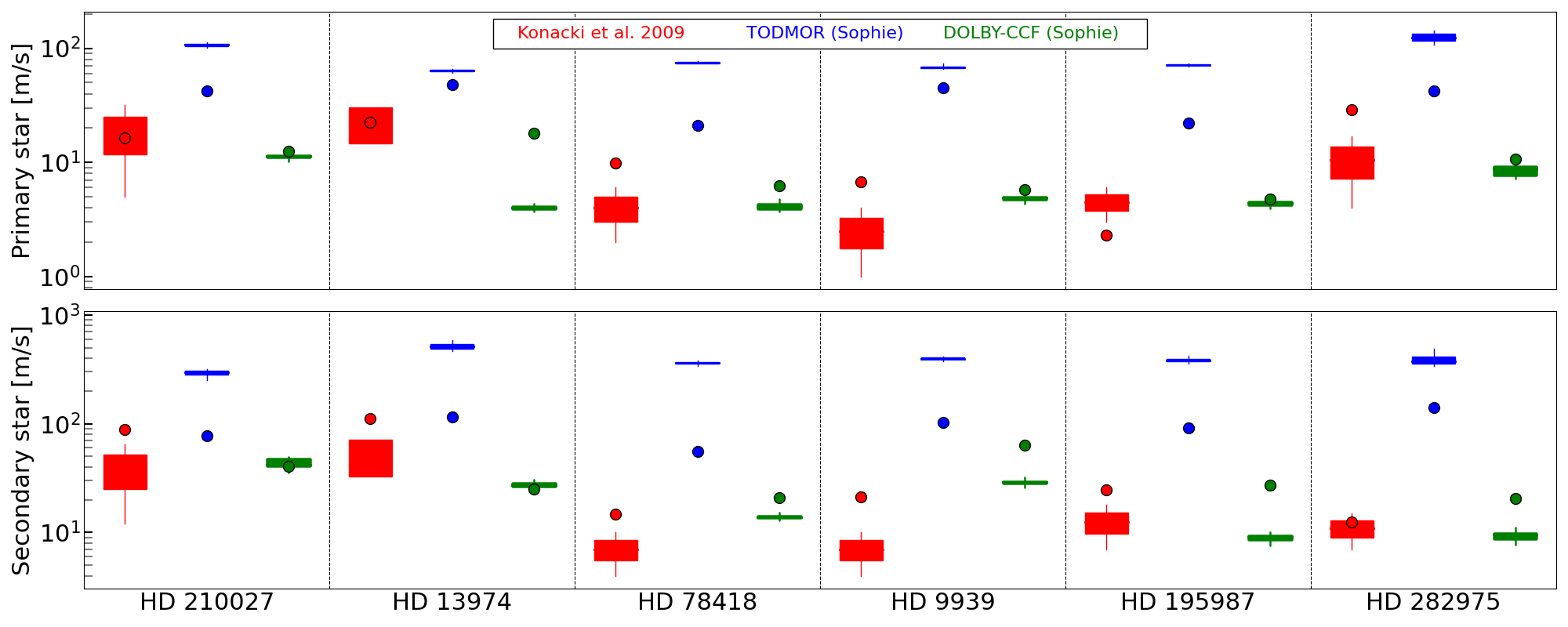}
\caption{The uncertainties are presented for three measurement methods: \citealt{Konacki_2009} (Keck/HIRES) - shown in red, TODMOR (Sophie) - shown in blue, and {\tt DOLBY}-CCF (Sophie) - shown in green. Box plots represent the distribution of uncertainties associated with each method for both primary (top panel) and secondary (bottom panel) stars. Additionally, RMS values for each method are plotted as scattered points with the corresponding colours. Note: For HD 13974, data from the Shane/CAT/Hamspec instrument (as reported in \citealt{Konacki_2009}) were used, as Keck/HIRES observations were not available for this system.}

\label{fig:uncertainity}
\end{figure*}

During our analysis, we attempted to compare the RMS values obtained from our {\tt kima}-{\tt DOLBY} analysis with those reported by \citep{Konacki_2009} for all our target systems. In Figure~\ref{fig:uncertainity}, the uncertainties are shown, where the box plot shows the distribution of uncertainties and the circles indicate the RMS for each binary system and compares them to other results. We focused on Keck/HIRES data for comparison, as this instrument provided the highest-precision measurements used by \citet{Konacki_2009}. For the case of HD 13974, as it was not observed with Keck/HIRES, we used the numbers from the Shane/CAT/Hamspec dataset (see Table 1 in \citealt{Konacki_2009}), which was the only available data.

Upon re-analysing the \citep{Konacki2010} RVs with kima, we found that RMS values were consistently higher than those reported. This discrepancy prevented a meaningful comparison between the results and also questioning accuracy of the RMS values. 

Although this re-analysis was carried out for just HD~78418 and HD~210027 in detail, the consistent overestimation of RMS values suggests that similar discrepancies might exist for other systems as well.

\section{Detection limits}

\begin{figure*}
    \centering
    \includegraphics[width=0.49\linewidth]{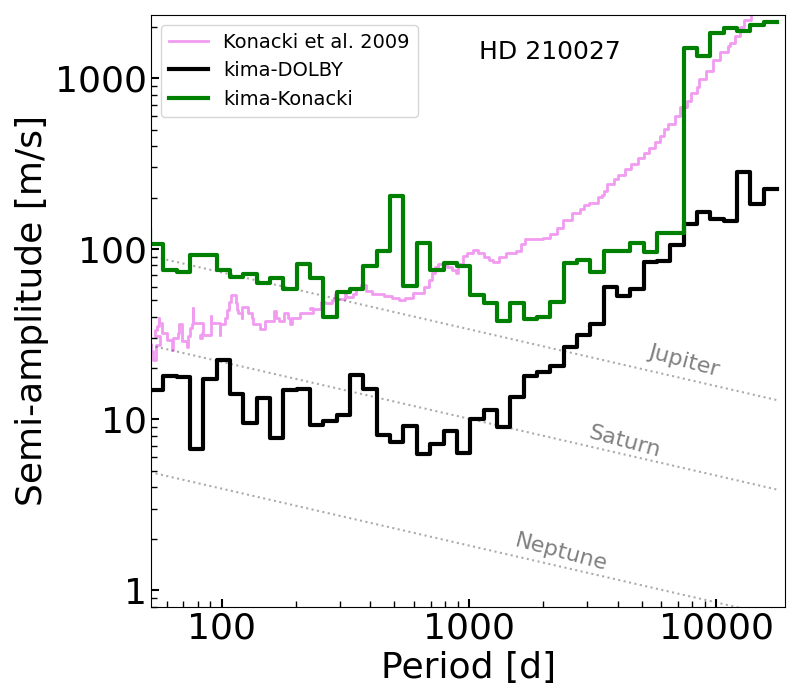}
    \includegraphics[width=0.49\linewidth]{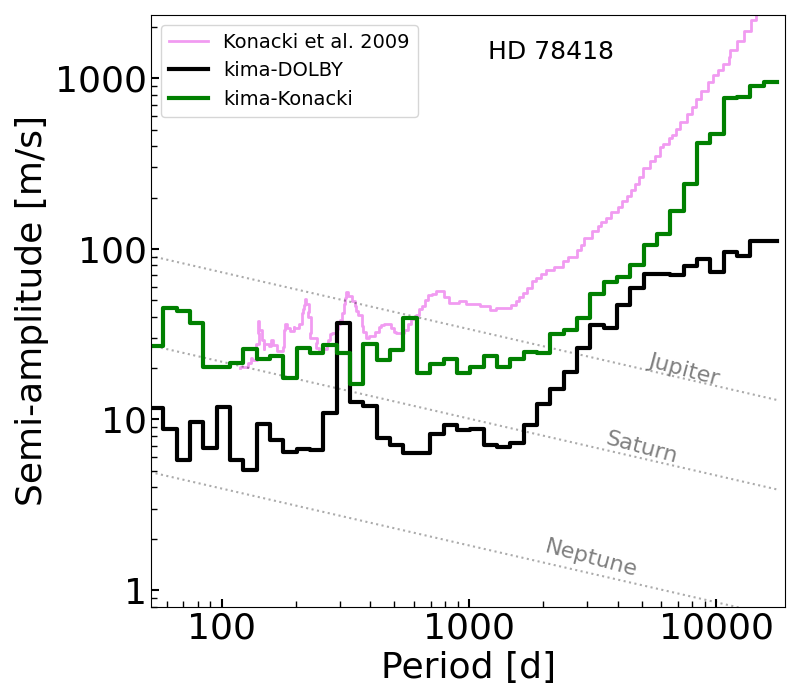}
    \caption{Detection limits based on the density of posterior samples for the HD~210027 and HD~78418  binary systems from the {\tt kima} run with N$_P$ fixed to 1. Solid lines represent the amplitude at the 99th percentile within each period bin, providing a 99\% confidence upper limit. The black line indicates the combined dataset analysis, while the green line shows the analysis using the {\tt kima}-Konacki derived data. The violet curve represents the planet detection limit from \citet{Konacki_2009}. Diagonal grey lines mark the expected signals for Neptune, Saturn, and Jupiter mass planets orbiting stars with their respective masses.}
    \label{fig:kima_compare}
\end{figure*}

When comparing the detection limits of the Konacki data with {\tt kima} (we call {\tt kima}-Konacki) and with their own analysis (we call \citealt{Konacki_2009}), we find that while they are comparable in terms of their overall significance, they exhibit different shapes. In Figure \ref{fig:kima_compare}, we show the comparison of these detection limits for HD~78418 and HD~210027. Both sets of detection limits using \cite{Konacki_2009} data are significantly above the detection limits obtained from our {\tt kima}-{\tt DOLBY} analysis, indicating that our {\tt kima}-{\tt DOLBY} method is more sensitive in detecting potential exoplanets.

\begin{figure*}
    \centering
    \includegraphics[width=0.95\linewidth]{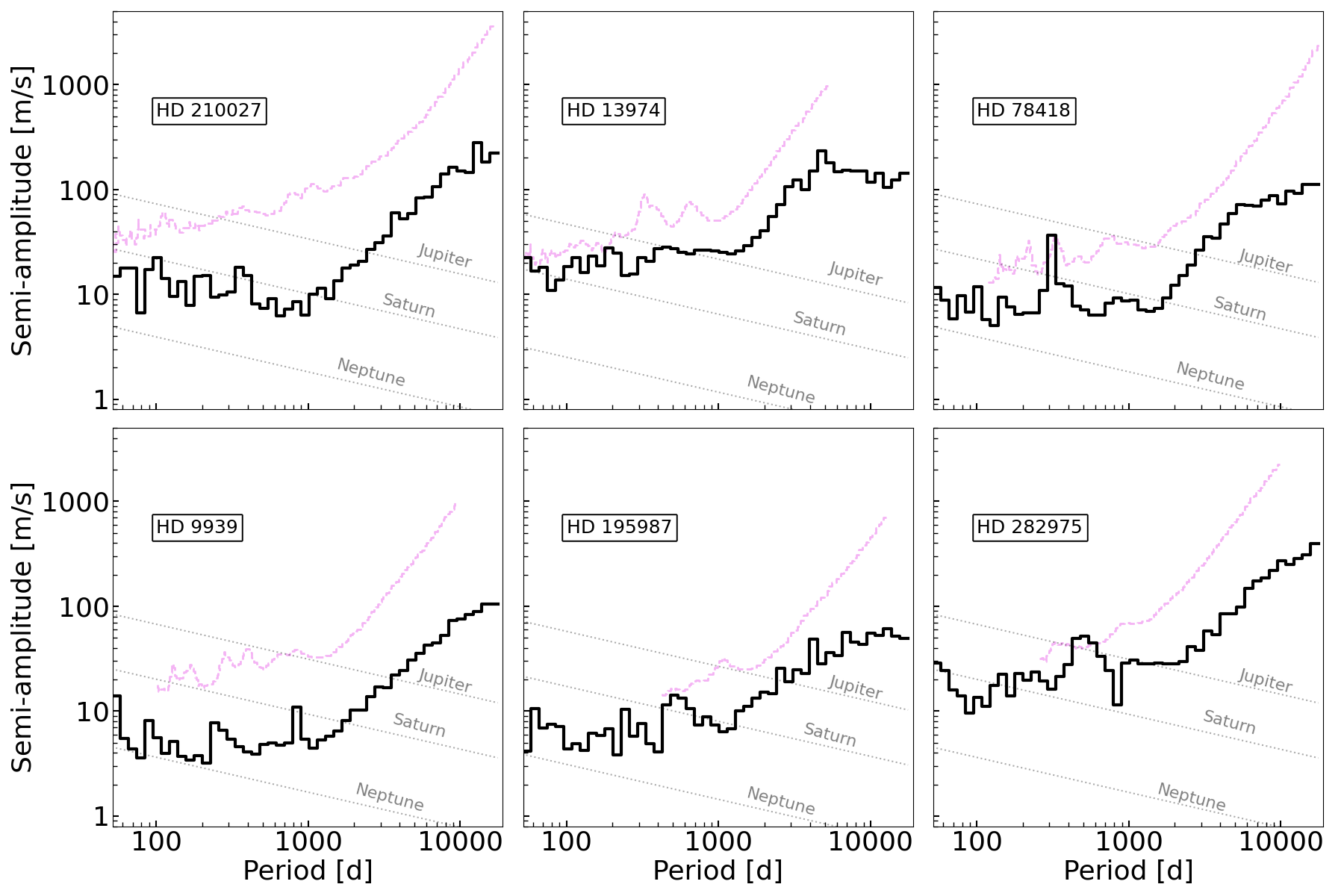}
    \caption{Detection limits for binary systems from the {\tt kima} run with N$_{\mathrm{P}}$ fixed to 1. Histograms represent the distribution of semi-amplitudes within each period bin, with the vertical lines indicating the 99th percentile, corresponding to a 99\% confidence upper limit. The violet curves show the planet detection limits from \citet{Konacki_2009,Konacki2010}, scaled to the number of observations obtained for our paper, to enable as close to a like for like comparison as possible. The diagonal gray lines represent the expected signals of Neptune, Saturn, and Jupiter mass planets orbiting stars with their respective binary masses.}
    \label{fig:detection_limit_scaled}
\end{figure*}

In Figure~\ref{fig:detection_limit_scaled}, we scale the planet detection limits from \citet{Konacki_2009} to number of observations and compare with {\tt DOLBY} detection limits. Our method consistently performs better than previous approaches in detecting exoplanets. This improvement is due to both our DOLBY method's accuracy and the increased number of measurements. Even when accounting for the number of measurements, {\tt DOLBY} still demonstrates significantly better detection limits. This suggests that {\tt DOLBY}'s underlying methodology is a key factor in our improved performance.

\section{Radial velocity data}

\begin{table*}
    \caption{RV Data for HD~210027 using Dolby CCF Method on Sophie Spectrum}
        \begin{tabular}{|c|c|c|c|c|c|c|}
        \hline
        BJD [days] & RV$_1$ [$\rm km\,s^{-1}$] & $\sigma_{\mathrm{RV}_1}$ [$\rm km\,s^{-1}$]& O-C$_1$ [$\rm km\,s^{-1}$]& RV$_2$ [$\rm km\,s^{-1}$] & $\sigma_{\mathrm{RV}_2}$ [$\rm km\,s^{-1}$] & O-C$_2$ [$\rm km\,s^{-1}$]\\
        \hline
2459123.38091 & -52.34232 & 0.01335 & 0.00938 & 72.11175 & 0.06193 & -0.01038 \\ 
2459123.38979 & -52.29537 & 0.01307 & 0.00453 & 72.01824 & 0.06096 & -0.02087 \\ 
2459128.45234 & 42.93887 & 0.01271 & 0.01924 & -80.69803 & 0.06467 & 0.01439 \\ 
2459128.48099 & 42.75522 & 0.01251 & -0.00456 & -80.47051 & 0.06449 & -0.01475 \\ 
2459132.51401 & -47.98416 & 0.01216 & 0.01493 & 65.10774 & 0.04950 & -0.03731 \\ 
2459132.51798 & -48.05310 & 0.01206 & -0.00057 & 65.17262 & 0.04864 & -0.05809 \\ 
2459137.39725 & 35.19619 & 0.01284 & -0.00349 & -68.35485 & 0.06139 & -0.03636 \\ 
2459154.37782 & -49.12542 & 0.01119 & 0.03364 & 66.98514 & 0.04214 & -0.01948 \\ 
2459158.44617 & 42.58362 & 0.01174 & 0.02644 & -80.14782 & 0.03846 & -0.01739 \\ 
2459162.39265 & -33.48042 & 0.01064 & 0.01004 & 41.85057 & 0.04570 & -0.03419 \\ 
2459168.33550 & 39.52619 & 0.01138 & -0.02196 & -75.30244 & 0.04575 & -0.00299 \\ 
2459170.35326 & 28.16629 & 0.01078 & -0.00049 & -57.08909 & 0.03863 & -0.05967 \\ 
2459174.31727 & -52.94527 & 0.01137 & -0.00282 & 73.04889 & 0.04186 & -0.02013 \\ 
2459176.35188 & -14.50687 & 0.01179 & -0.02429 & 11.31581 & 0.04117 & -0.08710 \\ 
2459181.33461 & 8.41498 & 0.01145 & -0.01973 & -25.32817 & 0.03894 & 0.03443 \\ 
2459185.26110 & -45.73935 & 0.01163 & -0.00722 & 61.52951 & 0.03888 & 0.01826 \\ 
2459351.62279 & 30.55845 & 0.01095 & 0.01034 & -60.89566 & 0.03765 & -0.04413 \\ 
2459354.61166 & 17.93299 & 0.01050 & 0.00754 & -40.54485 & 0.04092 & 0.04753 \\ 
2459374.58515 & 28.81861 & 0.01051 & 0.02387 & -58.06133 & 0.03695 & -0.02396 \\ 
2459381.58875 & 14.16772 & 0.01027 & 0.02032 & -34.50062 & 0.03904 & 0.02841 \\ 
2459389.60043 & -44.41779 & 0.01143 & 0.01247 & 59.43247 & 0.04067 & 0.00821 \\ 
2459398.56787 & -52.70122 & 0.01133 & -0.01393 & 72.66380 & 0.04248 & 0.00383 \\ 
2459410.59366 & -32.47557 & 0.01028 & -0.00835 & 40.18931 & 0.04127 & -0.05512 \\ 
2459414.44339 & 42.76507 & 0.01115 & -0.01156 & -80.55705 & 0.04341 & -0.07423 \\ 
2459424.49613 & 43.46540 & 0.01114 & -0.00656 & -81.63438 & 0.04170 & -0.03518 \\ 
2459425.57194 & 30.45091 & 0.01052 & 0.01262 & -60.67250 & 0.03548 & 0.00302 \\ 
2459426.60554 & 3.87457 & 0.01105 & 0.00766 & -18.00283 & 0.04177 & 0.03073 \\ 
2459427.49154 & -22.03691 & 0.01100 & -0.00689 & 23.44005 & 0.04240 & -0.06716 \\ 
2459429.57455 & -53.12680 & 0.01122 & 0.00713 & 73.33900 & 0.04205 & -0.03692 \\ 
2459433.45656 & 33.09481 & 0.01060 & 0.01271 & -64.87356 & 0.04046 & 0.04559 \\ 
2459463.49870 & 19.65339 & 0.01058 & 0.01002 & -43.36343 & 0.03972 & -0.01442 \\ 
2459489.37951 & -37.38137 & 0.01114 & 0.00679 & 48.10777 & 0.04249 & -0.02636 \\ 
2459503.26544 & -11.62182 & 0.01189 & 0.02698 & 6.83203 & 0.04118 & -0.02556 \\ 
2459514.40352 & 15.37954 & 0.01102 & -0.00493 & -36.48393 & 0.03982 & 0.03028 \\ 
2459520.30192 & -43.10972 & 0.01134 & -0.00755 & 57.31461 & 0.04368 & 0.01968 \\ 
2459538.32512 & 21.05451 & 0.01098 & 0.00717 & -45.67957 & 0.03954 & -0.07695 \\ 
2459548.24825 & 27.88614 & 0.01079 & -0.01125 & -56.68909 & 0.03870 & -0.09207 \\ 
2459553.30195 & -38.75507 & 0.01120 & 0.02067 & 50.28388 & 0.03966 & -0.07523 \\ 
2459560.22768 & -21.10140 & 0.01147 & -0.00414 & 21.94564 & 0.04147 & -0.06566 \\ 
2459565.27406 & 9.95154 & 0.01123 & -0.00173 & -27.81866 & 0.04056 & -0.01984 \\ 
2459741.57939 & 40.04336 & 0.01126 & 0.00540 & -76.05172 & 0.04490 & 0.03423 \\ 
2459774.60436 & -18.34614 & 0.01109 & 0.00123 & 17.59785 & 0.04022 & -0.00323 \\ 
2459776.61296 & -53.18037 & 0.02209 & 0.00901 & 73.39141 & 0.04257 & -0.07335 \\ 
2459797.43653 & -52.36365 & 0.01130 & 0.01846 & 72.23316 & 0.04337 & 0.06229 \\ 
2459853.40173 & 43.56933 & 0.01094 & -0.00846 & -81.80592 & 0.04184 & -0.03683 \\ 
2459866.25707 & -10.62429 & 0.01320 & 0.02584 & 5.27034 & 0.04887 & 0.01512 \\ 
2459866.26082 & -10.75328 & 0.01319 & 0.00745 & 5.47359 & 0.04757 & 0.04095 \\ 
2459867.45408 & -41.37190 & 0.01131 & 0.00510 & 54.49047 & 0.04287 & -0.03873 \\ 
2459888.30601 & -48.34365 & 0.01142 & 0.00867 & 65.64090 & 0.04064 & -0.07038 \\ 
2459900.31845 & -43.26153 & 0.01145 & 0.00560 & 57.59416 & 0.04333 & 0.03451 \\ 
2459924.23470 & 41.02977 & 0.01158 & 0.01571 & -77.67567 & 0.04813 & -0.02275 \\ 
2459924.30571 & 41.67963 & 0.01133 & 0.01377 & -78.76368 & 0.04836 & -0.06429 \\ 
2459924.30953 & 41.70923 & 0.01134 & 0.01082 & -78.76205 & 0.04827 & -0.01040 \\ 
2459924.31335 & 41.73751 & 0.01140 & 0.00679 & -78.83872 & 0.04868 & -0.03522 \\ 
2459924.31714 & 41.76263 & 0.01127 & 0.00013 & -78.85060 & 0.04891 & 0.00394 \\ 
2459924.32097 & 41.80818 & 0.01133 & 0.01381 & -78.96153 & 0.04902 & -0.05582 \\       
\hline
    \end{tabular}
\label{tab:HD210027_rvs}
\end{table*}

\begin{table*}
    \caption{RV Data for HD~13974 using Dolby CCF Method on Sophie Spectrum}
    \begin{tabular}{|c|c|c|c|c|c|c|}
        \hline
        BJD [days] & RV$_1$ [$\rm km\,s^{-1}$] & $\sigma_{\mathrm{RV}_1}$ [$\rm km\,s^{-1}$]& O-C$_1$ [$\rm km\,s^{-1}$]& RV$_2$ [$\rm km\,s^{-1}$] & $\sigma_{\mathrm{RV}_2}$ [$\rm km\,s^{-1}$] & O-C$_2$ [$\rm km\,s^{-1}$]\\
        \hline
2458851.36986 & -1.73619 & 0.00467 & -0.00665 & -13.04684 & 0.03064 & -0.01498 \\ 
2458860.32662 & 2.80378 & 0.00399 & -0.00540 & -19.38319 & 0.02867 & -0.05570 \\ 
2458871.33474 & -1.32472 & 0.00887 & -0.01535 & -13.66542 & 0.02769 & -0.05077 \\ 
2458895.26721 & -16.07826 & 0.00392 & -0.00223 & 6.84277 & 0.02696 & -0.02227 \\ 
2459068.59190 & -0.72497 & 0.00391 & 0.01157 & -14.41216 & 0.04891 & -0.00296 \\ 
2459094.56982 & -15.89868 & 0.00386 & -0.01229 & 6.59235 & 0.02681 & -0.00972 \\ 
2459100.65292 & 3.13285 & 0.00405 & -0.03291 & -19.81134 & 0.02858 & 0.01077 \\ 
2459104.66467 & -16.06569 & 0.00388 & -0.02251 & 6.80556 & 0.02724 & -0.01393 \\ 
2459123.57672 & -11.74338 & 0.00401 & 0.01705 & 0.89295 & 0.02872 & 0.01266 \\ 
2459133.53153 & -11.39425 & 0.00434 & 0.01315 & 0.45675 & 0.03071 & 0.06606 \\ 
2459136.54978 & -13.45844 & 0.00373 & 0.00943 & 3.28183 & 0.02582 & 0.03366 \\ 
2459138.58370 & -1.51518 & 0.00431 & 0.03473 & -13.28270 & 0.03927 & -0.00169 \\ 
2459158.53279 & -2.05556 & 0.00539 & 0.01251 & -12.57175 & 0.04432 & -0.00944 \\ 
2459164.52865 & -15.41809 & 0.00382 & 0.00156 & 5.91695 & 0.02634 & -0.03787 \\ 
2459165.48529 & -16.47060 & 0.00399 & 0.02934 & 7.49353 & 0.02685 & 0.04064 \\ 
2459173.53048 & -10.93347 & 0.00542 & 0.02043 & -0.22219 & 0.03723 & 0.01605 \\ 
2459174.57244 & -15.48454 & 0.00384 & 0.00199 & 6.02874 & 0.02634 & -0.01881 \\ 
2459180.33942 & 3.66827 & 0.00408 & 0.02940 & -20.48049 & 0.02669 & -0.00209 \\ 
2459184.46913 & -15.10267 & 0.00382 & 0.01293 & 5.54808 & 0.02630 & 0.01491 \\ 
2459190.33616 & 3.66605 & 0.00410 & 0.02769 & -20.47263 & 0.02809 & 0.00505 \\ 
2459200.25548 & 3.65633 & 0.00411 & 0.04599 & -20.44373 & 0.02720 & -0.00491 \\ 
2459244.27223 & -13.92165 & 0.00373 & 0.01211 & 3.89420 & 0.02600 & -0.00005 \\ 
2459249.27867 & 0.81736 & 0.00393 & -0.02331 & -16.54747 & 0.02595 & 0.04944 \\ 
2459415.60524 & -16.48543 & 0.00400 & -0.00738 & 7.41897 & 0.02737 & -0.00356 \\ 
2459459.57600 & 0.25367 & 0.00506 & -0.00209 & -15.77196 & 0.03292 & 0.01363 \\ 
2459489.63908 & 0.24514 & 0.00387 & -0.02267 & -15.79874 & 0.02523 & 0.00356 \\ 
2459501.49659 & 3.08327 & 0.00404 & -0.00942 & -19.72562 & 0.02812 & -0.00485 \\ 
2459542.39883 & 0.16392 & 0.00384 & -0.01913 & -15.71181 & 0.02459 & -0.02705 \\ 
2459622.27606 & 1.42672 & 0.00393 & -0.01359 & -17.43741 & 0.02701 & -0.00872 \\ 
2459775.60673 & -14.94069 & 0.00399 & -0.00263 & 5.27494 & 0.02393 & -0.01203 \\ 
2459832.51807 & 2.12060 & 0.00397 & 0.00688 & -18.34861 & 0.02794 & 0.01418 \\ 
2459855.59138 & -14.28766 & 0.00379 & 0.00169 & 4.38260 & 0.02654 & -0.00477 \\ 
2459870.56860 & 1.00991 & 0.00412 & 0.00156 & -16.80347 & 0.02777 & 0.02604 \\ 
2459921.44118 & 3.32644 & 0.00407 & -0.00759 & -20.07147 & 0.02768 & -0.01594 \\ 
2459958.40976 & -12.50915 & 0.00371 & -0.01640 & 1.87455 & 0.02592 & -0.02132 \\
        \hline
    \end{tabular}
\label{tab:HD13974_rvs}
\end{table*}

\begin{table*}
    \caption{RV Data for HD~78418 using Dolby CCF Method on Sophie Spectrum}
    \begin{tabular}{|c|c|c|c|c|c|c|}
        \hline
        BJD [days] & RV$_1$ [$\rm km\,s^{-1}$] & $\sigma_{\mathrm{RV}_1}$ [$\rm km\,s^{-1}$]& O-C$_1$ [$\rm km\,s^{-1}$]& RV$_2$ [$\rm km\,s^{-1}$] & $\sigma_{\mathrm{RV}_2}$ [$\rm km\,s^{-1}$] & O-C$_2$ [$\rm km\,s^{-1}$]\\
        \hline
2458872.37870 & 21.27676 & 0.00478 & 0.00243 & -3.59807 & 0.01528 & -0.02680 \\ 
2458898.49292 & -13.47233 & 0.00378 & 0.00228 & 36.76581 & 0.01414 & -0.01753 \\ 
2458907.47357 & 37.24499 & 0.00416 & 0.00233 & -22.09402 & 0.01499 & -0.02255 \\ 
2458910.41886 & 25.84510 & 0.00446 & 0.00470 & -8.88600 & 0.01388 & -0.01146 \\ 
2458911.33579 & 20.48118 & 0.00424 & -0.00490 & -2.68233 & 0.01424 & -0.02657 \\ 
2458920.37977 & -14.04810 & 0.00372 & 0.00328 & 37.47144 & 0.01403 & -0.01192 \\ 
2459130.66567 & -10.87384 & 0.00391 & 0.00873 & 33.76031 & 0.01343 & -0.01297 \\ 
2459153.68550 & -12.63554 & 0.00369 & 0.00408 & 35.81088 & 0.01367 & -0.00611 \\ 
2459157.62496 & 28.01406 & 0.00432 & 0.00501 & -11.42754 & 0.01325 & -0.04023 \\ 
2459172.58339 & -14.51336 & 0.00372 & 0.00670 & 38.01254 & 0.01380 & 0.01307 \\ 
2459181.62800 & 28.94156 & 0.00416 & 0.00581 & -12.46418 & 0.01292 & 0.00555 \\ 
2459186.60880 & -0.01687 & 0.00412 & -0.00104 & 21.14648 & 0.01415 & -0.03093 \\ 
2459237.62907 & 37.11821 & 0.00416 & 0.00216 & -21.93676 & 0.01489 & 0.03397 \\ 
2459248.49694 & -14.84864 & 0.00402 & -0.00381 & 38.38285 & 0.01461 & 0.00819 \\ 
2459248.63740 & -15.08212 & 0.00381 & 0.00540 & 38.66170 & 0.01424 & 0.00514 \\ 
2459263.45847 & 4.40171 & 0.00510 & 0.00464 & 16.01318 & 0.01736 & 0.01311 \\ 
2459280.45118 & 18.80057 & 0.00418 & -0.00563 & -0.66523 & 0.01464 & 0.03949 \\ 
2459298.35123 & 27.60645 & 0.00439 & 0.00968 & -10.91746 & 0.01347 & -0.00294 \\ 
2459303.36350 & -1.53055 & 0.00416 & 0.00713 & 22.93540 & 0.01345 & 0.01393 \\ 
2459536.71790 & -3.63984 & 0.00433 & 0.00221 & 25.31719 & 0.01362 & -0.00330 \\ 
2459537.69503 & -8.28468 & 0.00422 & 0.00437 & 30.72184 & 0.01352 & -0.01972 \\ 
2459540.71828 & -15.61997 & 0.00388 & -0.00113 & 39.25905 & 0.01431 & -0.01525 \\ 
2459546.72643 & 34.37384 & 0.00399 & 0.00767 & -18.79004 & 0.01386 & -0.01617 \\ 
2459548.72050 & 36.20188 & 0.00411 & 0.00708 & -20.89273 & 0.01468 & 0.00826 \\ 
2459594.52691 & -1.41021 & 0.00420 & 0.01081 & 22.78949 & 0.01379 & 0.00348 \\ 
2459608.48351 & 30.07109 & 0.00402 & 0.00787 & -13.75660 & 0.01311 & 0.02268 \\ 
2459623.43923 & 27.20333 & 0.00442 & 0.00343 & -10.48153 & 0.01340 & -0.03430 \\ 
2459638.50612 & -14.43425 & 0.00382 & 0.01078 & 37.92500 & 0.01414 & 0.01261 \\ 
2459644.38323 & 36.64494 & 0.00421 & 0.00544 & -21.37130 & 0.01505 & 0.04458 \\ 
2459648.51053 & 23.43053 & 0.00443 & 0.01470 & -6.04911 & 0.01292 & 0.00938 \\ 
2459868.64759 & -12.01896 & 0.01200 & -0.01003 & 35.03915 & 0.04037 & 0.01023 \\ 
2459888.64574 & -13.80355 & 0.00371 & -0.00410 & 37.16529 & 0.01406 & 0.00471 \\ 
2459957.53755 & 31.86693 & 0.00397 & -0.00411 & -15.93369 & 0.01387 & -0.01783 \\ 
2459958.61724 & 26.23676 & 0.00441 & -0.00113 & -9.37006 & 0.01354 & -0.03385 \\ 
2459959.41779 & 21.59474 & 0.00434 & 0.00075 & -3.97159 & 0.01377 & -0.02906 \\ 
2460006.40723 & -15.65450 & 0.00392 & 0.00856 & 39.29607 & 0.01421 & 0.00431 \\ 
2460007.41061 & -14.22850 & 0.00376 & 0.01059 & 37.69088 & 0.01387 & 0.01750 \\
    \hline
    \end{tabular}
\label{tab:HD78418_rvs}
\end{table*}

\begin{table*}
    \caption{RV Data for HD~9939 using Dolby CCF Method on Sophie Spectrum}
    \begin{tabular}{|c|c|c|c|c|c|c|}
        \hline
        BJD [days] & RV$_1$ [$\rm km\,s^{-1}$] & $\sigma_{\mathrm{RV}_1}$ [$\rm km\,s^{-1}$]& O-C$_1$ [$\rm km\,s^{-1}$]& RV$_2$ [$\rm km\,s^{-1}$] & $\sigma_{\mathrm{RV}_2}$ [$\rm km\,s^{-1}$] & O-C$_2$ [$\rm km\,s^{-1}$]\\
        \hline
2458858.37230 & -69.59819 & 0.00515 & 0.00270 & -15.63101 & 0.03004 & -0.05248 \\ 
2459068.58880 & -55.01951 & 0.00495 & -0.00328 & -34.21645 & 0.02903 & -0.00851 \\ 
2459077.63446 & -17.38670 & 0.00501 & 0.00615 & -82.06566 & 0.03015 & 0.00370 \\ 
2459096.55610 & -27.41804 & 0.00492 & 0.00596 & -69.30955 & 0.02952 & 0.00033 \\ 
2459100.64992 & -8.86494 & 0.00498 & 0.00891 & -92.88379 & 0.02824 & -0.01363 \\ 
2459102.62882 & -16.15512 & 0.00490 & -0.00481 & -83.63979 & 0.02759 & 0.01531 \\ 
2459104.61855 & -29.54584 & 0.00513 & -0.00086 & -66.60837 & 0.03219 & 0.00029 \\ 
2459123.55891 & -13.54038 & 0.00483 & 0.00498 & -86.98169 & 0.02947 & 0.00025 \\ 
2459136.58872 & -73.58189 & 0.00507 & 0.00056 & -10.64629 & 0.03260 & -0.01308 \\ 
2459138.58018 & -78.44756 & 0.00496 & 0.00302 & -4.48104 & 0.03072 & 0.01384 \\ 
2459164.41903 & -78.74896 & 0.00489 & -0.00195 & -4.10826 & 0.03023 & -0.01774 \\ 
2459165.48006 & -77.71594 & 0.00511 & 0.00895 & -5.40001 & 0.03057 & 0.00166 \\ 
2459166.49085 & -74.77060 & 0.00505 & 0.00970 & -9.13958 & 0.03140 & -0.00145 \\ 
2459168.38319 & -63.83368 & 0.00511 & -0.00559 & -23.05031 & 0.03003 & -0.03120 \\ 
2459173.47593 & -16.64036 & 0.00493 & -0.00323 & -83.01520 & 0.02886 & 0.02846 \\ 
2459173.52642 & -16.29119 & 0.00499 & 0.00384 & -83.46716 & 0.02892 & 0.02492 \\ 
2459181.43455 & -38.62125 & 0.00508 & 0.00235 & -55.08504 & 0.02845 & -0.00032 \\ 
2459185.33203 & -65.65455 & 0.00515 & 0.00756 & -20.73494 & 0.03146 & -0.00352 \\ 
2459190.32281 & -78.30099 & 0.00493 & 0.00660 & -4.67014 & 0.03023 & 0.02450 \\ 
2459242.29567 & -74.04520 & 0.00514 & 0.00174 & -10.09967 & 0.03274 & 0.02638 \\ 
2459244.24760 & -61.97739 & 0.00524 & 0.00606 & -25.41548 & 0.03208 & 0.00685 \\ 
2459248.33469 & -22.62320 & 0.00504 & 0.00028 & -75.38557 & 0.03036 & 0.01407 \\ 
2459399.59509 & -22.57071 & 0.00497 & -0.00231 & -75.47639 & 0.02970 & -0.01187 \\ 
2459415.60256 & -78.12005 & 0.00502 & 0.00291 & -4.88995 & 0.03065 & 0.01973 \\ 
2459432.57414 & -31.36037 & 0.00599 & -0.00282 & -64.31752 & 0.03795 & 0.00535 \\ 
2459462.56743 & -65.31155 & 0.00512 & 0.00724 & -21.18090 & 0.03105 & -0.01767 \\ 
2459463.56289 & -70.45211 & 0.00518 & -0.01031 & -14.63985 & 0.02922 & -0.01885 \\ 
2459489.64627 & -74.07999 & 0.00512 & -0.00204 & -10.02058 & 0.03289 & -0.01304 \\ 
2459503.46598 & -8.53155 & 0.00502 & -0.00012 & -93.33199 & 0.02811 & 0.02007 \\ 
2459538.55434 & -67.29364 & 0.00513 & -0.00023 & -18.64480 & 0.03261 & -0.03109 \\ 
2459542.39462 & -78.74315 & 0.00494 & 0.00154 & -4.10112 & 0.03110 & 0.00871 \\ 
2459545.46940 & -70.75187 & 0.00515 & 0.00295 & -14.24031 & 0.02945 & 0.01263 \\ 
2459546.41498 & -64.59449 & 0.00514 & 0.00243 & -22.06312 & 0.03049 & 0.00750 \\ 
2459551.40013 & -18.13658 & 0.00515 & 0.00577 & -81.10746 & 0.03288 & -0.01380 \\ 
2459558.43304 & -29.96531 & 0.00520 & -0.00159 & -66.06794 & 0.03387 & 0.00023 \\ 
2459601.29306 & -22.33338 & 0.00497 & -0.00309 & -75.75891 & 0.02978 & -0.02485 \\ 
2459779.61956 & -10.65818 & 0.00492 & 0.00909 & -90.54018 & 0.03178 & -0.02012 \\ 
2459797.62825 & -70.34967 & 0.00519 & 0.00081 & -14.64547 & 0.03007 & -0.02225 \\ 
2459854.52788 & -14.02566 & 0.00486 & 0.00269 & -86.19184 & 0.02820 & -0.07637 \\ 
2459869.42980 & -78.21860 & 0.00496 & 0.00456 & -4.61381 & 0.03050 & -0.04957 \\ 
2459886.56099 & -33.09126 & 0.00509 & -0.00906 & -61.95479 & 0.02905 & -0.03289 \\ 
2459904.48732 & -16.94018 & 0.00501 & 0.01999 & -82.44737 & 0.02987 & -0.11632 \\ 
2459959.27294 & -13.77033 & 0.00482 & 0.00285 & -86.42047 & 0.02921 & 0.29986 \\
\hline
    \end{tabular}
\label{tab:HD9939_rvs}
\end{table*}

\begin{table*}
    \centering
    \caption{RV Data for HD~195987 using Dolby CCF Method on Sophie Spectrum}
    \begin{tabular}{|c|c|c|c|c|c|c|}
        \hline
        BJD [days] & RV$_1$ [$\rm km\,s^{-1}$] & $\sigma_{\mathrm{RV}_1}$ [$\rm km\,s^{-1}$]& O-C$_1$ [$\rm km\,s^{-1}$]& RV$_2$ [$\rm km\,s^{-1}$] & $\sigma_{\mathrm{RV}_2}$ [$\rm km\,s^{-1}$] & O-C$_2$ [$\rm km\,s^{-1}$]\\
        \hline
2458767.44980 & -19.58729 & 0.00347 & -0.00068 & 12.15788 & 0.00770 & 0.00767 \\ 
2459010.57241 & -24.99781 & 0.00519 & -0.00263 & 19.01268 & 0.00888 & -0.01264 \\ 
2459031.58433 & 24.52511 & 0.00415 & 0.00087 & -43.97312 & 0.00784 & -0.04054 \\ 
2459033.60485 & 30.51372 & 0.00442 & 0.00342 & -51.56721 & 0.01024 & -0.01998 \\ 
2459034.55544 & 31.80618 & 0.00506 & -0.00414 & -53.22300 & 0.01027 & -0.02187 \\ 
2459044.54125 & 1.75428 & 0.00399 & -0.00375 & -14.96329 & 0.01010 & 0.02017 \\ 
2459068.48549 & -24.73894 & 0.00355 & 0.00059 & 18.70550 & 0.00891 & 0.00512 \\ 
2459072.53538 & -21.77383 & 0.00582 & -0.00152 & 14.97086 & 0.01324 & 0.04196 \\ 
2459077.47004 & -14.56690 & 0.00376 & 0.00261 & 5.79191 & 0.00899 & 0.01842 \\ 
2459093.48616 & 31.30272 & 0.00377 & 0.00578 & -52.50469 & 0.00909 & 0.04346 \\ 
2459100.50645 & 6.67546 & 0.00598 & 0.00272 & -21.18494 & 0.01203 & 0.04708 \\ 
2459122.46379 & -25.65979 & 0.00465 & -0.00675 & 19.86971 & 0.00706 & 0.00826 \\ 
2459122.48541 & -25.65150 & 0.00417 & -0.00033 & 19.87494 & 0.00747 & 0.01586 \\ 
2459127.40713 & -23.83828 & 0.00342 & -0.00694 & 17.55151 & 0.00698 & 0.00547 \\ 
2459128.44366 & -23.06215 & 0.00526 & 0.00214 & 16.53802 & 0.01249 & -0.03307 \\ 
2459132.26642 & -18.82948 & 0.00611 & 0.00454 & 11.16976 & 0.01254 & -0.02435 \\ 
2459155.30948 & 16.79037 & 0.00418 & 0.00138 & -34.06186 & 0.00883 & 0.03363 \\ 
2459165.33216 & -14.33204 & 0.00283 & -0.00104 & 5.51959 & 0.00822 & 0.04999 \\ 
2459172.35297 & -23.24635 & 0.00510 & -0.00492 & 16.83323 & 0.00677 & 0.03730 \\ 
2459174.27477 & -24.46398 & 0.00422 & 0.00084 & 18.34732 & 0.00755 & -0.00367 \\ 
2459305.64369 & -16.61477 & 0.00566 & 0.00146 & 8.35839 & 0.01098 & -0.01670 \\ 
2459351.61913 & -25.66858 & 0.00439 & -0.00534 & 19.85431 & 0.01154 & -0.02010 \\ 
2459354.60816 & -24.94729 & 0.00360 & 0.00534 & 18.96805 & 0.00897 & -0.00318 \\ 
2459363.54500 & -15.57905 & 0.00500 & 0.00688 & 7.04920 & 0.01023 & -0.01627 \\ 
2459371.59079 & 8.05734 & 0.00568 & 0.00399 & -23.00347 & 0.01043 & -0.01700 \\ 
2459374.60037 & 20.87997 & 0.00439 & 0.00052 & -39.28299 & 0.01012 & 0.01382 \\ 
2459377.52102 & 30.48196 & 0.00470 & 0.00299 & -51.55269 & 0.00998 & -0.04533 \\ 
2459399.44040 & -21.26427 & 0.00414 & 0.00068 & 14.28020 & 0.00838 & -0.00339 \\ 
2459410.58249 & -25.39247 & 0.00386 & 0.00577 & 19.48908 & 0.00917 & -0.04853 \\ 
2459415.50033 & -22.68478 & 0.00408 & 0.00184 & 16.07992 & 0.00904 & -0.01112 \\ 
2459429.57920 & 10.84287 & 0.00366 & 0.00714 & -26.49598 & 0.00956 & 0.02835 \\ 
2459432.56152 & 23.46212 & 0.00445 & -0.00028 & -42.61899 & 0.00980 & -0.03699 \\ 
2459436.49438 & 32.01894 & 0.00414 & 0.00667 & -53.45443 & 0.00861 & 0.00369 \\ 
2459463.48393 & -25.43569 & 0.00453 & 0.00459 & 19.55807 & 0.00679 & -0.03288 \\ 
2459477.48547 & -16.82898 & 0.00492 & -0.00224 & 8.66785 & 0.00728 & 0.02519 \\ 
2459544.32381 & 11.25872 & 0.00382 & -0.00440 & -27.09618 & 0.00771 & -0.02841 \\ 
2459546.35246 & 19.97966 & 0.00529 & -0.00447 & -38.19812 & 0.01023 & -0.04000 \\ 
2459554.28903 & 25.65379 & 0.00382 & -0.00800 & -45.42088 & 0.00755 & -0.04084 \\ 
2459646.69694 & -20.74314 & 0.00373 & -0.00458 & 13.65525 & 0.00663 & 0.04032 \\ 
2459647.70492 & -19.47515 & 0.00380 & -0.00519 & 11.97384 & 0.00762 & -0.02860 \\ 
2459739.49580 & -16.01978 & 0.00545 & -0.00318 & 7.59803 & 0.01082 & -0.01423 \\ 
2459774.57537 & 15.41881 & 0.00458 & 0.00781 & -32.33024 & 0.00907 & 0.01204 \\ 
2459776.61934 & 23.95970 & 0.00447 & 0.00989 & -43.21266 & 0.00908 & -0.01070 \\ 
2459800.63752 & -21.21510 & 0.00434 & -0.00839 & 14.18785 & 0.00926 & -0.02171 \\ 
2459855.36671 & -17.94399 & 0.00422 & 0.00763 & 10.04320 & 0.01014 & -0.02873 \\ 
2459866.26356 & -25.66229 & 0.00430 & 0.00409 & 19.84599 & 0.01060 & -0.03239 \\ 
2459869.34471 & -25.34610 & 0.00370 & 0.00232 & 19.45066 & 0.00956 & -0.02363 \\ 
2459901.24851 & 13.68083 & 0.00463 & -0.00370 & -30.15247 & 0.01161 & -0.00477 \\ 
2459924.21804 & -25.68638 & 0.00385 & -0.00038 & 19.86468 & 0.01016 & -0.03864 \\ 
2460010.70041 & 31.20871 & 0.00500 & -0.00157 & -52.41145 & 0.01030 & 0.02645 \\ 
2460041.66411 & -25.23418 & 0.00358 & 0.00975 & 19.34025 & 0.00947 & -0.00124 \\ 
2460070.62334 & 24.11113 & 0.00473 & 0.01144 & -43.42312 & 0.00977 & -0.02995 \\
\hline
    \end{tabular}
\label{tab:HD195987_rvs}
\end{table*}

\begin{table*}
    \caption{RV Data for HD~282975 using Dolby CCF Method on Sophie Spectrum}
    \begin{tabular}{|c|c|c|c|c|c|c|}
        \hline
        BJD [days] & RV$_1$ [$\rm km\,s^{-1}$] & $\sigma_{\mathrm{RV}_1}$ [$\rm km\,s^{-1}$]& O-C$_1$ [$\rm km\,s^{-1}$]& RV$_2$ [$\rm km\,s^{-1}$] & $\sigma_{\mathrm{RV}_2}$ [$\rm km\,s^{-1}$] & O-C$_2$ [$\rm km\,s^{-1}$]\\
        \hline
2458851.38469 & 0.45294 & 0.00978 & -0.00576 & 14.09135 & 0.01029 & -0.03460 \\ 
2458898.38654 & -0.37863 & 0.00897 & 0.00240 & 15.00823 & 0.00983 & -0.00228 \\ 
2458924.31871 & -0.20532 & 0.00965 & 0.00559 & 14.81763 & 0.01026 & -0.01379 \\ 
2459096.59172 & 12.95871 & 0.00923 & 0.01624 & 1.00437 & 0.00954 & 0.02252 \\ 
2459122.61784 & 12.91231 & 0.00945 & -0.01052 & 0.98820 & 0.00975 & -0.01433 \\ 
2459136.62423 & -0.67026 & 0.00729 & 0.00498 & 15.33625 & 0.00768 & 0.01638 \\ 
2459184.50882 & 0.26131 & 0.00932 & 0.01278 & 14.34905 & 0.00934 & 0.00129 \\ 
2459241.40466 & -0.17399 & 0.00768 & 0.00414 & 14.81492 & 0.00863 & 0.01846 \\ 
2459475.58271 & -0.39070 & 0.00717 & -0.00602 & 14.99645 & 0.00822 & -0.01749 \\ 
2459513.54933 & 13.15543 & 0.00809 & 0.00320 & 0.74992 & 0.00929 & -0.01107 \\ 
2459514.55909 & 14.01542 & 0.00750 & 0.00432 & -0.13764 & 0.00858 & 0.00567 \\ 
2459544.64636 & 12.94248 & 0.00833 & 0.00038 & 0.99053 & 0.00936 & 0.00777 \\ 
2459554.47041 & 0.30082 & 0.00970 & 0.00159 & 14.27782 & 0.01111 & -0.01603 \\ 
2459569.38751 & 14.52205 & 0.00735 & 0.00098 & -0.67618 & 0.00821 & 0.00387 \\ 
2459642.30919 & 11.64299 & 0.01070 & 0.01542 & 2.36058 & 0.01226 & -0.00574 \\ 
2459801.61970 & 14.38294 & 0.00916 & 0.00460 & -0.52525 & 0.01032 & 0.00472 \\ 
2459854.62748 & 14.76487 & 0.00732 & 0.01723 & -0.87450 & 0.00808 & 0.04427 \\ 
2459866.54498 & -0.13297 & 0.00848 & 0.01645 & 14.78591 & 0.00906 & 0.01968 \\ 
2459903.52873 & 12.42914 & 0.00872 & -0.01281 & 1.47036 & 0.00980 & -0.03849 \\ 
2459955.38046 & 12.18273 & 0.00876 & -0.00584 & 1.74739 & 0.00981 & -0.02825 \\ 
2459958.48351 & 14.63086 & 0.00886 & -0.02523 & -0.84999 & 0.00968 & -0.02759 \\
    \hline
    \end{tabular}
\label{tab:HD282975_rvs}
\end{table*}

%%%%%%%%%%%%%%%%%%%%%%%%%%%%%%%%%%%%%%%%%%%%%%%%%%

% Don't change these lines
%\bsp	% typesetting comment
\label{lastpage}
\end{document}